\documentclass[]{elsarticle}

\usepackage{amsmath,amssymb,amsthm,amsfonts,mathrsfs}
\newtheorem{thm}{Theorem}
\newtheorem{lem}[thm]{Lemma}
\newdefinition{rmk}{Remark}
\newproof{pf}{Proof}
\newproof{pot}{Proof of Theorem \ref{thm2}}
\usepackage[colorlinks,linkcolor=blue,anchorcolor=green,citecolor=blue]{hyperref}%%xlpeng
\usepackage{natbib}

\journal{Elsevier}

\begin{document}

\begin{frontmatter}

%\title{The network structure and SIV epidemic model in random walkers}
\title{An SIS epidemic model with vaccination in a dynamical contact network of mobile individuals with heterogeneous spatial constraints}

\author[1,2]{Xiao-Long Peng\corref{cor1}}
\ead{xlpeng@sxu.edu.cn}

\author[1,2,3]{Ze-Qiong Zhang}

\author[1,2]{Junyuan Yang}

\author[1,2]{Zhen Jin}

\cortext[cor1]{Corresponding author}

\address[1]{Complex Systems Research Center, Shanxi University, Taiyuan 030006, Shanxi, China}
\address[2]{Shanxi Key Laboratory of Mathematical Techniques and Big Data Analysis on Disease Control and Prevention, Shanxi University, Taiyuan 030006, Shanxi, China}
\address[3]{School of Mathematical Sciences, Shanxi University, Taiyuan 030006, Shanxi, China}

\begin{abstract}
Network-based epidemic models have been extensively employed to understand the spread of infectious diseases, but have generally overlooked the fact that most realistic networks are dynamical rather than static. In this paper, we study a susceptible-infected-susceptible epidemic model with vaccination in a dynamical contact network of moving individuals, where we regard mobile individuals as random walkers that are allowed to perform long-range jumps. Different from previous studies of epidemics in a random walk network with a constant interaction radius, we consider the scenario where the individuals have a heterogeneous distribution of interaction radius $r$ and infected individuals are vaccinated with a probability depending on the interaction radius distribution. We derive the basic reproduction number $\mathcal{R}_0$, which is strongly related to the interaction radius distribution and is proportional to the second order moment of interaction radius $\langle r^2\rangle$ in the special case of a constant vaccination rate. We argue that if $\mathcal{R}_0<1$ then the disease-free equilibrium is locally asymptotically stable, whereas if $\mathcal{R}_0>1$ then there is a unique endemic equilibrium which is locally asymptotically stable and uniformly persistent. In addition, we provide a sufficient condition for the global stability of the disease-free equilibrium. Both theoretical and simulation results reveal that the distribution of individual interaction radius has significant effects on the basic reproduction number and the final epidemic prevalence. In general, the disease will break out more readily in the population with a more heterogeneous radius distribution, while it will end in a lower epidemic prevalence. Interestingly, the results suggest that an optimal vaccination intervention for disease prevention and control is achievable regardless of the radius distribution. Furthermore, some interesting results on the structure of the underlying contact network are shown to have strong correlation with the epidemic dynamics. This study provides potential implications for developing efficient containment measures against infectious disease while considering the spatial constraints of moving individuals.
\end{abstract}

\begin{keyword}
\texttt{Dynamical network}\sep Epidemic spreading\sep Vaccination \sep Spatial constraint
\PACS 89.75.Hc\sep 89.75.-k \sep 87.23.Ge
\end{keyword}

\end{frontmatter}

%\linenumbers

\section{Introduction}

%\paragraph{Installation} If the document class \emph{elsarticle} is not available on your computer, you can download and install the system package \emph{texlive-publishers} (Linux) or install the \LaTeX\ package \emph{elsarticle} using the package manager of your \TeX\ installation, which is typically \TeX\ Live or Mik\TeX.

As yet, a powerful technique to study infectious diseases spreading in populations has been to build a mathematical compartment model \cite{May92,Hethcote00,Keeling08,Martcheva15} through which we can grasp how the disease evolves in time and provide implications for devising effective disease control measures. Two classical examples, among others, are the susceptible-infected-susceptible (SIS) model in which individuals can be reinfected after recovery and the susceptible-infected-recovered (SIR) model in which individuals gain lifelong immunity after infection \cite{May92,Hethcote00,Keeling08,Martcheva15}. Though very useful for dynamical analysis, these compartment models nevertheless have heavily hinged upon the homogeneous mixing (HM) assumption that all the individuals have the same probability to contact each other \cite{May92,Hethcote00,Keeling08,Martcheva15}. However, the contact pattern among individuals in real populations is heterogeneous such that the number of contacts or interactions varies widely among individuals \cite{Barabasi99,Liljeros01}. In recent decades, complex networks \cite{Barabasi99, Watts98,Newman03siam,Boccaletti06} have become a repeatedly used paradigm to study the spread of infectious diseases since the network-based epidemiological models \cite{Keeling05,Pastor15,Jin16,WangTang17,Xuxj17,Wuq18,Sunm18,Chens18,Zhangx18} have transcended the limitation of HM assumption in compartment models. The underlying network structures have been found to entail significant impacts on epidemic spreading dynamics \cite{Pastor15} and other dynamical processes taking place over networks \cite{Barrat08}. As a well-known example, Pastor-Satorras and Vespignani \cite{Pastor01} discovered that the epidemic threshold for the SIS epidemic model vanishes in the thermodynamic limit in a static scale-free (SF) network \cite{Barabasi99} which obeys a power-law degree distribution $P(k)\propto k^{-\gamma}$ with an exponent $2<\gamma\leq 3$. It is worth remarking that a more mathematically rigorous analysis showed that the critical value is also zero for contact processes on random networks with power-law degree distributions for any value of power $\gamma>3$ \cite{Chatterjee09}.

A large body of investigations on network epidemiology have focused on static networks, where the topological structure of network is fixed in time during the process of disease transmission \cite{Fu08,Peng13,Wu14,ZhangHF14,ZhuGH15,WangZ16,WuZX16,WangY17,ZhangHF17,Ding18}, thus ignoring the impact of motions of individuals. In reality, individuals in a population often move around during the spread of infectious diseases, leading to structural changes of the underlying contact network that mimics the population \cite{Riley07,Brockmann13}. Recently, the epidemic spreading on random walk networks \cite{Buscarino14,Frasca06,Buscarino10,Xia10,Zhou09,Yang12,Li15,Gan12} has been extensively investigated to understand the effects of individuals' motion on the epidemic dynamics. For instance, Frasca \textit{et al.} \cite{Frasca06} proposed a dynamical network model with mobile individuals who are allowed to perform both local and long-distance motions. In their model, mobile individuals are modeled as random walkers who are only able to interact with others falling within a given interaction radius apart from them. Based on a similar dynamical network model, Buscarino \textit{et al.} \cite{Buscarino10} argued that the homogeneous mixing approximation is appropriate only when the velocity of individuals' movement is large enough.

On the one hand, most infectious disease models presented in the literature have largely neglected the influence of spatial distances between individuals. In fact, many realistic networks such as the mobile phone communication network, social contact network and the power grid are often embedded in a Euclidean geographical space \cite{Barthelemy11} and the interactions among individuals usually depend on their spatial distances and geographical information \cite{Rozenfeld02}. Generally speaking, the living space and the sphere of activity of individuals are constrained in terms of spatial distances \cite{Rattana14,Broder-Rodgers15}. Moreover, because of individual diversity, the activity ability, activity range, contact number and the geographic location \cite{Simini12} differ from individual to individual. Typically the limitation of the individuals on the spatial distance will lead to the localization of the contact pattern between individuals, which will affect the transmission of infectious diseases throughout the contact network \cite{Rattana14,Broder-Rodgers15}. Therefore, it is natural to study the infectious diseases model in contact networks with geographical properties, such as the embedded lattice \cite{Rozenfeld02} and the spatially embedded networks \cite{Emmerich14}. Xu $et$ $al.$ presented an SIS epidemic model in a lattice-embedded scale-free network and investigated how the geographical structure affects the dynamical process of epidemic spreading \cite{Xu06}. As a further step, Xu $et$ $al.$ also considered the standard SIS model on a random growing network to study the integrated effects of preference and geography on epidemic spreading \cite{Xu07}. In such spatially embedded networks, the individuals' interaction radius is generally assumed to be primarily determined by their respective degrees, that is, the larger degree, the larger interaction radius \cite{Rozenfeld02,Emmerich14,Xu06,Xu07}. It is worth remarking that most previous mentioned works based on random walk networks \cite{Buscarino14,Frasca06,Buscarino10,Xia10} have simply assumed that all individuals have the same interaction radius in order to better include other factors such as the velocity and the direction of motion, as well as the population density. However, the interaction radius of individuals in realistic populations or networks are usually heterogeneous \cite{Huang16}. For example, individuals with poor personal hygiene are prone to have a larger radius of contacting infectious sources. In a wireless sensor communication network, sensors with different power have different communication radii \cite{Shakkottai05}. Most of the aforementioned works concentrated on the threshold analysis of the model under consideration. However, there is little (if any) work devoted to stability analysis on epidemic models in complex networks with spatial or geographical constraints in the literature. In this paper, we consider an epidemic model with heterogeneous interaction radius of individuals, based on which we derive the basic reproduction number, analyze the equilibria stability, prove the model persistence, and investigate the effects of spatial constraints in individuals' mobilities and vaccination intervention on the epidemic spreading and on the network structure.

On the other hand, vaccination is one of the most effective policies for preventing the transmission of infectious diseases \cite{Bauch03,Althouse10} and up to now there have been a large number of studies on various vaccination strategies for epidemic models in complex networks \cite{Takeuchi06,Shaban08,LiCui09,Rushmore14,Huang17,ZhangShu17}. Traditional vaccination methods include random and targeted immunization strategies \cite{Pastor02}. It is argued that random immunization strategy is insufficient for networks with broad degree distributions, whereas targeted vaccination is to immunize high-degree nodes and has a much higher effectiveness than random vaccination in SF networks \cite{Pastor02} and small-world networks \cite{Zanette02}. However, it is difficult to implement the targeted vaccination strategy since it requires full knowledge of the degree of each node in the network. To overcome this shortcoming, several vaccination strategies based on local information have been proposed, such as acquaintance immunization \cite{Cohen03}. All these mentioned vaccination strategies are based on the degree distribution of individuals of the network. Rather than relying on node degree, in this paper we adopt individuals' interaction radius to characterize the individual heterogeneity and propose a new vaccination strategy that depends on the interaction radius of each individual.

The contributions of this paper are as follows. An SIS epidemic model with interaction radius-dependent vaccination is proposed to probe the potential effects of heterogeneous spatial constraints of individuals on epidemic spreading in a dynamic contact network of moving individuals. The epidemic dynamics is described by a set of ordinary differential equations. The explicit mathematical expression of the basic reproduction number is derived and the dynamical properties of both the disease-free equilibrium and the endemic equilibrium of the model system are presented. Numerical calculation and stochastic simulation show good agreement, indicating that our model can well describe the dynamical process of disease transmission on dynamical networks of mobile individuals. The effects of different radius distributions on epidemic dynamics and network structure have been examined. Our results imply that the optimal vaccination intervention is realizable for disease prevention and control.

This paper is outlined as follows. In section \ref{section2}, we describe the construction of a random walk network in which individuals are allowed to perform long-distance jumps with a probability and then present the SIS epidemic model with vaccination that depends on individuals' interaction radius. In section \ref{section3} we derive the basic reproduction number $\mathcal{R}_0$. We give the stability analysis of equilibria in Section \ref{section4} and discuss the persistence of the model in Section \ref{section5}. In section \ref{section6} we give the simulation results and discussion. In section \ref{section7}, we conclude the paper.

\section{Model description} \label{section2}

\subsection{Dynamic contact network of mobile individuals}

\begin{figure}[htp]
\centering
\includegraphics[scale=0.65]{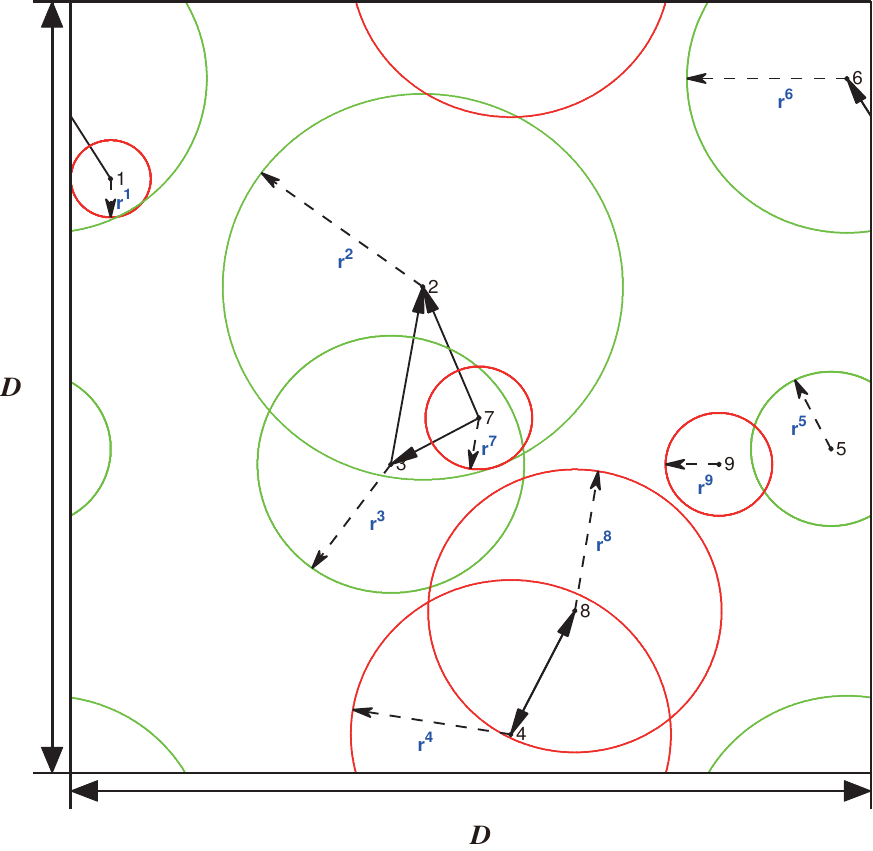}
\caption{(Color online) Diagrammatic sketch for epidemic spreading in a dynamical contact network (random walk network) of moving agents in a $D\times D$ planar space with periodic boundary conditions and spatial constraints. In this diagram there are nine individuals (nodes) with heterogeneous interaction radius $r^j$, $j=1,\dots,9$, indicated by dashed lines. Green and red circles represent the effective interaction (or contact) range of susceptible and infected individuals, respectively. Each susceptible individual $i$ can only be infected by infected individuals who run into the green circle (with the interaction radius $r^i$) around the individual $i$. For example, as demonstrated by the solid lines with arrows, the infected individual $1$ can infect the individual $6$, whilst the individual $9$ can not infect the individual $5$. Both of the individuals $2$ and $3$ can be infected by the individual $7$.}\label{fig1}
\end{figure}

There are a number of works that rely on random walk network to inspect the effects of moving agents on epidemic spreading \cite{Buscarino14,Frasca06,Buscarino10,Yang12,Huang16}. In a similar framework, we consider $N$ individuals who are initially randomly distributed in a two-dimensional space $\Omega=\{{(x, y)}\in\mathbb{R}^2:0\le x\le D, 0\le y\le D\}$, with periodic boundary conditions, as illustrated in Fig.~\ref{fig1}.
For convenience, we denote $\Lambda_i(t)=(x_i(t), y_i(t))$ as the position of the individual $i$ $(i=1,2,\dots,N)$ in the planar space with its moving velocity $v_i(t)=(v\cos\eta_i(t), v\sin\eta_i(t))$ and moving direction $\eta_i(t)$  at time $t$, where $v$ is the modulus of the agent velocity, which is the same for all individuals. Then, the motion of individual $i$ can be described as follows:
\begin{equation*}
\begin{array}{ll}
\displaystyle  x_i(t+1)=x_i(t)+v\cos\eta_i(t),  \\
\displaystyle  y_i(t+1)=y_i(t)+v\sin\eta_i(t),  \\
\displaystyle  \eta_i(t+1)=\varepsilon_i(t+1),  \\
\end{array}
\end{equation*}
where $\varepsilon_i(t)$ is a random variable obeying the uniform distribution between the interval $[-\pi,\pi]$. In addition, to include the probability that individuals can move with time scales much shorter than those related to disease, we consider the case where infected individuals may perform long-distance jumps. We define a parameter $p_{\rm{jump}}$ that quantifies the probability for an individual to perform a long-distance jump. Each individual can jump to any position (i.e. long-distance jump) inside the planar space with the probability $p_{\rm{jump}}$, that is, $p_{\rm{jump}}$ denotes the probability of an individual jumping to a random position in the space $\Omega$, similar to the case of other works \cite{Buscarino14,Frasca06,Buscarino10,Huang16}. Each individual evolves following $v_i(t)=(v\cos\eta_i(t), v\sin\eta_i(t))$ with probability $1-p_{\rm{jump}}$ or performs a random jump with probability $p_{\rm{jump}}$. In what follows, the model is investigated as a function of the parameter $p_{\rm{jump}}$. At time $t$, the Euclidean distance between individual $i$ and $j$ is defined as
\begin{equation*}
d_{ij}(t)=d_{ji}(t)=\sqrt{\big(x_i(t)-x_j(t)\big)^2+\big(y_i(t)-y_j(t)\big)^2}, \quad i,j=1,2,\dots,N.
\end{equation*}

\begin{figure}[htp]
\centering
\includegraphics[scale=0.8]{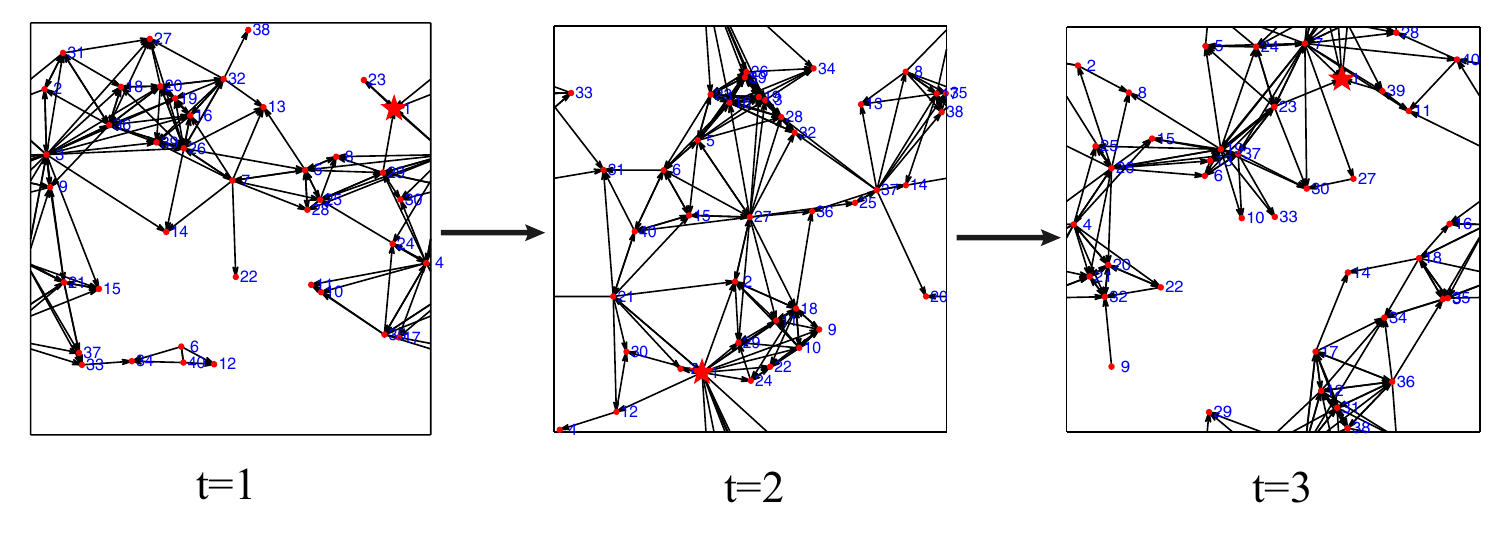}
\caption{(Color online) Schematic illustration of the dynamic contact network of 40 mobile individuals with spatial constraints. Each individual in the network displaces randomly at each time step. The position of individual $1$ at times $t=1,2,3$ has been marked by the red pentagram. The arrowed line from individual $i$ pointing to individual $j$ indicates that individual $i$ drops into the radius of individual $j$. The random jump probability is set to be $p_{\text jump}=1$.}\label{fig2}
\end{figure}

In the present work, we consider that each individual's behavior is constrained by their respective spatial distance, which is characterized by the interaction radius $r$ in the following. In order to take account of individual heterogeneity, we consider an individual $i$ has its own interaction radius $r^i$, i=1,2,\dots,N. Here, the interaction radius is the effective interaction distance on the Euclidean plane, denoting the characteristic radius of the circular region within which individuals can get infections from others. In our model, we assume there are $m$ different values of interaction radius which obey a preassigned probability distribution $P(r_j)$, $j=1,\dots,m$, where $P(r_j)$ denotes the proportion of nodes with interaction radius $r_j$. Namely, the interaction radius of each individual is given once and for all. At any time $t$, individual $i$ can only be able to interact with other individuals $j (\neq i)$ that fall within the circle defined by individual $i$'s position $\Lambda_i(t)$ and its interaction radius $r^i$. In the context of disease transmission, individual $i$ can be infected by any infected individuals who are located within the circle defined by the location $\Lambda_i$ and interaction radius $r^i$ of individual $i$. In this sense, the interaction radius can be seen as the ``susceptibility'' radius \cite{Huang16}. As demonstrated in Fig.~\ref{fig1}, it is possible for the disease to spread from the individual $1$ to individual $6$, whereas it is impossible to spread from individual $9$ to individual $5$ since the individual $9$ does not enter the realm of individual $5$. The definition of interaction radius forms a dynamical directed contact network \cite{Meyers06,Bernhardsson06}, as illustrated in Fig.~\ref{fig2}. All the individuals comprise the nodes of the contact network, in which the contacts are defined asymmetrically such that a node $j$ is regarded as a effective contact or a neighbor (which is capable of disease transmission) of node $i$ only if node $j$ is located in the circular realm of node $i$, but the converse is not necessarily true. In the layout given in Fig.~\ref{fig1}, for example, both the nodes $3$ and $7$ are neighbors of node $2$, while the node $2$ is neither the neighbor of node $3$ nor the neighbor of node $7$. Nodes $4$ and $8$ are mutual effective contacts to each other since either of them is positioned within the interaction radius of the other. Note that when all the individuals share an identical interaction radius, i.e., the values of interaction radius follow a delta distribution. The in-degree and out-degree of nodes of the directed network \cite{Meyers06,Bernhardsson06} can be defined as follows. The in-degree of node $i$ at time $t$ depends on the number of other nodes that fall into the interaction radius $r^i$ of node $i$. That is, the in-degree of node $i$ is defined as the number of effective contacts in the realm of node $i$. More clearly, the in-degree of node $i$ is the number of nodes $j$ who satisfy $d_{ij}(t)\leq r^i, \forall j\neq i$. Similarly, the out-degree of node $i$ is defined as the number of nodes whose effective contacts include node $i$, that is, the number of nodes $j$ who meet $d_{ij}(t)\leq r^j, \forall j\neq i$.

\subsection{SIS epidemic model with vaccination}

In this paper, we consider an SIS epidemic model with vaccination, where the vaccination of susceptible individuals depends on their interaction radius. The choice of such a vaccination strategy is motivated by the following consideration. We observe that a lot of studies on vaccination strategies take the node degree as the characteristic index of individual heterogeneity \cite{Fu08,Peng13,Pastor02,Cohen03,Gong13,Peng16,Wu1601,Wu1602,Chang17}. As an alternative measurement to quantify the diversity among individuals, interaction radius can characterize the range and ability of individuals' activity \cite{Huang16}. In this regard, we consider different vaccination rate for susceptible individuals according to their interaction radius.

\begin{figure}[htp]
\centering
\includegraphics[scale=0.8]{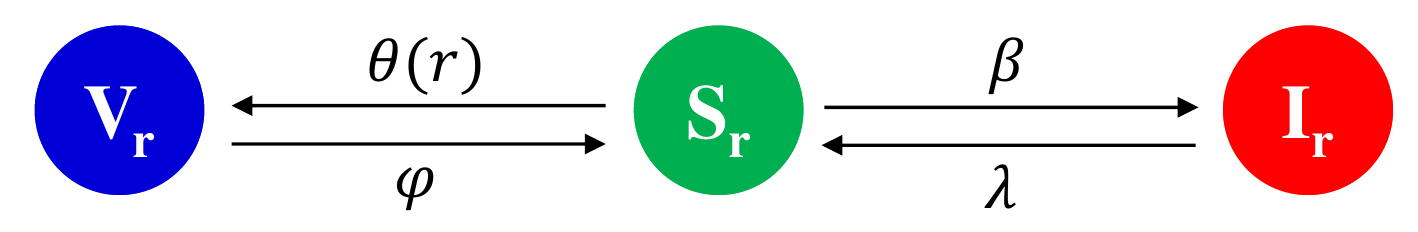}
\caption{(Color online) Schematic illustration of the SIS epidemic model with vaccination. Here, we use $S_r$, $I_r$ and $V_r$ to denote susceptible, infected and vaccinated nodes with effective interaction radius $r$. At each time step, each susceptible nodes is infected by infected neighbours at the transmission rate $\beta$ and is vaccinated at the vaccination rate $\theta(r)$, which relies on its interaction radius $r$. Infected nodes recover and return to being susceptible with the recovery rate $\lambda$. Vaccinated nodes become susceptible with the resusceptibility rate $\varphi$.}\label{fig3}
\end{figure}

In the model, there are a total number $N$ of individuals, each of which may have only one of the three possible states: susceptible (S), infected (I) and vaccinated (V). We denote $N_S(t)$, $N_I(t)$ and $N_V(t)$ as, respectively, the number of susceptible, infected and vaccinated individuals at time $t$. Obviously, we have the total number of individuals $N_S(t)+N_I(t)+N_V(t)=N$, which remains constant over time.

The epidemic spreading process follows the transmission rule as shown in Fig.~\ref{fig3}, where the individuals of states S, I and V are all subscripted with an interaction radius $r$. At each time step, each susceptible individual can be infected by an infected neighbor with the transmission rate $\beta$ and be vaccinated with the vaccination rate $\theta(r)$ that depends on the interaction radius of the susceptible individual. Each infected individual recovers at the recovery rate $\lambda$, and each vaccinated individual returns to being susceptible again with the resusceptibility rate $\varphi$ after the vaccine wears off.

Here, we consider the vaccination rate $\theta(r)$ in the form
\begin{equation}\label{func0}
\theta(r)=\theta_0\frac{P(r)r^\alpha}{\sum_{j=1}^{m} P({r_j}){r_j}^\alpha},
\end{equation}
where $\alpha\in \mathbb{R}$ is a tunable parameter and  $0<\theta_0<1$ is a constant. Equation~(\ref{func0}) implies that the vaccination of an individual depends on the individual's effective interaction radius $r$ and the probability $P(r)$ that the individual's interaction radius is $r$. When $\alpha=0$, it means that the vaccination rate depends only on the probability distribution $P(r)$ of the interaction radius. When $\alpha\neq0$, the vaccination rate is $\theta(r)\propto P(r)r^\alpha$, similar to the preferential attachment hypothesis in the growing network model \cite{Romualdo01}. The case of $\alpha>0$ means that the larger the value of the radius $r$ and its probability $P(r)$, the higher the vaccination rate; while the case of $\alpha<0$ indicates that the larger the value of $r$ and the smaller the probability $P(r)$, the smaller the vaccination rate. In the context of disease transmission, the case of $\alpha>0$ accounts for the scenario in which individuals who are active in social contacts have a higher risk of infections and should be vaccinated with high priority. On the contrary, the case of $\alpha<0$ considers the situation that individuals with close contacts should be protected preferentially.

Aiming to understand the system behavior of the epidemic spreading over the dynamical contact network of mobile individuals, we derive a mean-field model based on the HM assumption that the population mixes at random, i.e., all the individuals have the same probability to contact the other ones \cite{May92,Hethcote00,Keeling08,Martcheva15}. It has been argued that if the individuals' mobility is extremely high, namely, if $p_{\rm{jump}}\rightarrow 1$ or $v\simeq D$ is satisfied, the underlying contact network is degenerated into an averaged one that can be well approximated by the HM hypothesis \cite{Buscarino14,Frasca06,Buscarino10,Huang16}. In our model, the parameter $p_{\rm{jump}}$ quantifies the probability for an individual to perform a long-distance jump to a random position in the planar space. When $p_{\rm{jump}}=1$, all the individuals jump to a random position in the two-dimensional space independently. In this case, the spatial correlations in the disease states are destroyed by the agent motion \cite{Buscarino08}. Therefore, the case of $p_{\rm{jump}}=1$ is equivalent to the case where all the individuals have the same chance to encounter the others. In a word, the case of $p_{\rm{jump}}=1$ in our model can be well approximated by the HM assumption.

Under the HM hypothesis, we have the density of individuals $\rho=N/{D^2}$ which can be fixed by fixing the value of $D$. Taking into account the heterogeneous interaction radius of individuals, we denote the number of susceptible, infected and vaccinated individuals with effective interaction (susceptibility) radius $r$ by $N_S(r,t)$, $N_I(r,t)$ and $N_V(r,t)$, respectively. It is straightforward to get the number of individuals with effective interaction radius $r$ as $N_r=N_S(r,t)+N_I(r,t)+N_V(r,t)$ which is a constant only depending on the probability distribution $P(r)$. We further define $s(r,t)=N_S(r,t)/N_r$, $i(r,t)=N_I(r,t)/N_r$ and $v(r,t)=N_V(r,t)/N_r$ as the fraction (or relative density) of individuals with effective interaction radius $r$ at time $t$, respectively in the susceptible, infected and vaccinated states. Obviously, $s(r,t)+i(r,t)+v(r,t)=1$. The equations for the dynamical system of mobile individuals in different disease states read
\begin{equation}\label{func1}
\begin{array}{ll}
\displaystyle  \frac{ds(r,t)}{dt}=\lambda i(r,t)+\varphi v(r,t)-s(r,t)\theta(r)-s(r,t)[1-(1-\beta)^{k_r^{inf}}],\\
\displaystyle  \frac{di(r,t)}{dt}=-\lambda i(r,t)+s(r,t)[1-(1-\beta)^{k_r^{inf}}], \\
\displaystyle  \frac{dv(r,t)}{dt}=s(r,t)\theta(r)-\varphi v(r,t), \quad r=r_1,r_2,\dots,r_m.
\end{array}
\end{equation}
The first equation refers to creation of susceptibles because of recovery of infecteds (first term), increase of susceptibles due to resusceptibility of vaccinateds (second term), loss of susceptibles due to vaccination (third term) and decrease of susceptibles caused by infection that is proportional to $s(r,t)$ times a contagion probability $p_{\rm {cont}}$. The contagion probability is given by $p_{\rm{cont}}=1-\bar{p}_{\rm{cont}}$, where $\bar{p}_{\rm{cont}}$ is the probability of not being infected. Since $\bar{p}_{\rm{cont}}$ is the probability that an individual with radius $r$ is not infected by any of its infected neighbors at time $t$, we have $\bar{p}_{\rm{cont}}=(1-\beta)^{k_r^{inf}}$, where $k_r^{inf}$ is the number of its infected neighbors. Based on the mean-field approximation and the HM hypothesis, we have $k_r^{inf}=\rho\pi r^2\sum_{j=1}^m P(r_j)i(r_j,t)$ \cite{Huang16}. Therefore, the fraction of susceptible individuals with radius $r$ that enter the infected compartment at time $t+1$ is $s(r,t)[1-(1-\beta)^{k_r^{inf}}]$. The second equation indicates that the decrease of infecteds is proportional to the fraction of infecteds which get recovered, i.e., to $\lambda i(r,t)$, and that the increase of infecteds is proportional to the density of susceptibles contracting the disease. The third equation is derived by considering the vaccination of susceptibles with rate $\theta(r)$ and the relapse into susceptibility with rate $\varphi$ for vaccinated individuals as the vaccine wears off.

As the transmission rate $\beta$ is small enough, we can make the approximation $[1-(1-\beta)^{k_r^{inf}}]\approx\beta {k_r^{inf}}$. In this paper, the epidemiological parameters are set to be small since the choice of parameter values only affects the time scale of the disease propagation without influencing the generality of the results \cite{Xia10,ZhouT06,Zanette08}. Following such an approximation, the model system of Eqs.~(\ref{func1}) can be simplified to
\begin{equation}\label{func2}
\begin{array}{ll}
\displaystyle  \frac{ds(r,t)}{dt}=\lambda i(r,t)-s(r,t)\beta\rho\pi r^2\sum_{j=1}^m P(r_j)i(r_j,t)+\varphi v(r,t)-s(r,t)\theta(r) \\
\displaystyle  \frac{di(r,t)}{dt}=-\lambda i(r,t)+s(r,t)\beta\rho\pi r^2\sum_{j=1}^m P(r_j)i(r_j,t) \\
\displaystyle  \frac{dv(r,t)}{dt}=s(r,t)\theta(r)-\varphi v(r,t), \quad r=r_1,r_2,\dots,r_m.
\end{array}
\end{equation}

The initial conditions of the model system (\ref{func2}) take the form
\begin{equation}\label{init}
\begin{array}{ll}
s(r,0)+i(r,0)+v(r,0)=1,\\
0\leq s(r,0)\leq 1,\\
0\leq i(r,0)\leq 1,\\
0\leq v(r,0)\leq 1, \quad r=r_1,r_2,\dots,r_m.
\end{array}
\end{equation}
Combined with the initial conditions (\ref{init}) and the preassigned interaction radius distribution $P(r)$, the system (\ref{func2}) determines the epidemic dynamics on the spatial contact network of mobile individuals. In order to present the dynamical properties of the system, in what follows we provide a lemma on the positiveness and boundedness of solutions to system (\ref{func2}).

\begin{lem}\label{lem1}
Let $\big(s(r,t),i(r,t),v(r_,t)\big), r=r_1,r_2,\dots,r_m$ be the solutions to system (\ref{func2}) with the initial conditions given by (\ref{init}), then it follows that
\begin{equation*}
0\leq s(r,t),i(r,t),v(r,t) \leq 1, \quad s(r,t)+i(r,t)+v(r,t)=1
\end{equation*}
for any $r=r_1,r_2,\dots,r_m$ and $t\geq 0$.
\end{lem}

\begin{pf}
Firstly, we verify $i(r,t)\geq 0$ for any $r=r_1,r_2,\dots,r_m$. By the way of contradiction, because $i(r,0)\geq 0$, we assume that there exist some $r_0\in\{r_1,r_2,\dots,r_m\}$ and $t\geq 0$ such that $i(r_0,t)=0$. Let
\begin{equation*}
t_0=inf\{t\geq 0\big{|}i(r_0,t)=0\},
\end{equation*}
then $i(r_0,t_0)=0$, $\frac{di(r_0,t_0)}{dt}<0$ and $i(r_0,t)>0$ for any $t\in[0,t_0)$. It follows from the second equation of system (\ref{func2}) that
\begin{equation*}
\frac{di(r_0,t_0)}{dt}=s(r_0,t_0)\beta\rho\pi r_0^2{\sum_{j=1}^{m} P(r_j)i(r_j,t_0)}<0.
\end{equation*}
This indicates that $s(r_0,t_0)<0$. Since $s(r_0,0)\geq 0$, there exists a $t_1<t_0$ such that $s(r_0,t_1)=0$, $\frac{ds(r_0,t_1)}{dt}<0$ and $s(r_0,t)>0$ for any $t\in [0,t_1)$.

By the first equation of system (\ref{func2}), we have
\begin{equation*}
\frac{ds(r_0,t_1)}{dt}=\lambda i(r_0,t_1)+\varphi v(r_0,t_1)<0,
\end{equation*}
which implies that $v(r_0,t_1)<0$, since $i(r_0,t_1)>0$.

Similarly, there exists a $t_2<t_1$ such that $v(r_0,t_2)=0$, $\frac{dv(r_0,t_2)}{dt}<0$ and $v(r_0,t)>0$ for any $t\in [0,t_2)$. Substituting $v(r_0,t_2)=0$ into the last equation of system (\ref{func2}) yields
\begin{equation*}
\frac{dv(r_0,t_2)}{dt}=s(r_0,t_2)\theta(r_0)<0.
\end{equation*}
This means $s(r_0,t_2)<0$ which leads to a contradiction with $s(r_0,t)> 0$ for any $t\in [0,t_1)$. Hence, $i(r,t)\geq 0$ for any $r=r_1,r_2,\dots,r_m$ and any $t\geq 0$.

In a similar way, using the way of contradiction starting from the third equation of system (\ref{func2}), we can easily show that $v(r,t)\geq 0$ for any $r=r_1,r_2,\dots,r_m$ and any $t\geq 0$.

Now we assume there exist some $r_0\in\{r_1,r_2,\dots,r_m\}$ and $t\geq 0$ such that $s(r_0,t)=0$. Let
\begin{equation*}
t_0=inf\{t\geq 0\big{|}s(r_0,t)=0\},
\end{equation*}
then $s(r_0,t_0)=0$, $\frac{ds(r_0,t_0)}{dt}<0$ and $s(r_0,t)>0$ for any $t\in[0,t_0)$. It follows from the first equation of system (\ref{func2}) that
\begin{equation*}
\frac{ds(r_0,t_0)}{dt}=\lambda i(r_0,t_0)+\varphi v(r_0,t_0)<0,
\end{equation*}
which is a contraction since $i(r_0,t_0)\geq 0$ and $v(r_0,t_0)\geq 0$. This contradiction indicates that $s(r,t)\geq 0$ for any $r=r_1,r_2,\dots,r_m$ and any $t\geq 0$.

Therefore, it is straightforward to obtain $s(r,t)\leq 1$, $i(r,t)\leq 1$, and $v(r,t) \leq 1$ because $s(r,t)+i(r,t)+v(r,t)=1$. The proof of Lemma \ref{lem1} is completed. \qed

\end{pf}

\section{Equilibria and basic reproduction number}\label{section3}

In this section, we will derive the basic reproduction number $\mathcal{R}_0$ by examining the existence and uniqueness of the endemic equilibrium (EE) of our model. In mathematical epidemiology, the basic reproduction number is an important threshold indicator that determines whether the disease breaks out or dies out. It is defined as the average number of new infections caused by an infected individual during its infectious period when appearing in a completely susceptible population \cite{May92,Hethcote00,Keeling08,Martcheva15}. Generally, if $\mathcal{R}_0>1$ then the disease will break out resulting in an endemic state; otherwise if $\mathcal{R}_0<1$ the disease will become extinct eventually \cite{May92,Hethcote00,Keeling08,Martcheva15}.

Based on the normalization condition $s(r,t)+i(r,t)+v(r,t)=1$, the model (\ref{func2}) can be reduced to
\begin{equation}\label{func4}
\begin{array}{ll}
\displaystyle  \frac{ds(r,t)}{dt}=&\lambda i(r,t)-s(r,t)\beta\rho\pi r^2\sum_{j=1}^m P(r_j)i(r_j,t)-s(r,t)\theta(r) \\
&+\varphi\big[1-s(r,t)-i(r,t)\big],   \\
\displaystyle  \frac{di(r,t)}{dt}=&-\lambda i(r,t)+s(r,t)\beta\rho\pi r^2\sum_{j=1}^m P(r_j)i(r_j,t),\quad r=r_1,r_2,\dots,r_m.
\end{array}
\end{equation}

According to Lemma \ref{lem1}, the feasible region for system (\ref{func4}) is given by
\begin{equation}\label{func4b}
\begin{array}{ll}
\Gamma =&\Bigg\{\Big(s(r_1,t),i(r_1,t),\dots,s(r_m,t),i(r_m,t)\Big)\in \mathbb{R}^{2m}\Big{|} 0\leq s(r,t)\leq1,  \\
 &0\leq i(r,t)\leq 1, 0\leq s(r,t)+i(r,t)\leq1, r=r_1,r_2,\dots,r_m\Bigg\},
\end{array}
\end{equation}
which is positively invariant with regard to system (\ref{func4}).

Obviously, system (\ref{func4}) admits a unique disease-free equilibrium (DFE)
\begin{equation}\label{func5}
E_0=\Big(\frac{\varphi}{\varphi+\theta(r_1)},0,\frac{\varphi}{\varphi+\theta(r_2)},0,\dots,
\frac{\varphi}{\varphi+\theta(r_m)},0\Big),
\end{equation}
on the boundary $\partial\Gamma$ of the invariant set $\Gamma$.

By letting the right-hand side of (\ref{func4}) be zero, we have a stationary solution of system (\ref{func4}) in the limit of $t\rightarrow\infty$ as
\begin{equation}\label{func6}
\displaystyle  i_r^*=\frac{{\beta\varphi\rho\pi r^2}{\sum_{j=1}^m P(r_j)i_{r_j}^*}}{\lambda\big(\varphi+\theta(r)\big)+\beta\varphi\rho\pi r^2{\sum_{j=1}^m P(r_j)i_{r_j}^*}},
\end{equation}
where $i_r^*=\lim\limits_{t\rightarrow\infty} i(r,t)$, $r=r_1,r_2,\dots,r_m$.

Denote by $I(t)\in[0,1]$ the fraction of infected individuals among the total $N$ individuals and by $I^*$ the stationary value of $I(t)$ as $t\rightarrow\infty$. By this definition, we have
\begin{equation}\label{func7}
\displaystyle  I(t)=\sum_{j=1}^m P(r_j)i(r_j,t), \quad I^*=\sum_{j=1}^m P(r_j)i^*_{r_j}.
\end{equation}
Note that $0\leq i(r,t)\leq 1$ for any $r=r_1,r_2,\dots,r_m$ and $t\geq 0$ according to Lemma \ref{lem1}, thus we have $0\leq i_r^*\leq 1$ for any $r=r_1,r_2,\dots,r_m$, and hence $0\leq I^*\leq 1$.

Combining Eqs.~(\ref{func6}) and (\ref{func7}) gives rise to a self-consistency equation
\begin{equation}\label{func8}
\displaystyle  I^*=\sum_{j=1}^m P(r_j){\frac{\beta\varphi\rho\pi r_j^2 I^*}{\lambda\big(\varphi+\theta(r_j)\big)+\beta\varphi\rho\pi r_j^2 I^*}},
\end{equation}
which implies a trivial solution $I^*=0$. Now we give the conditions about the existence and uniqueness of the nontrivial positive solution $I^*>0$. To this aim, define
\begin{equation*}
\displaystyle  F(I^*)=\sum_{j=1}^m P(r_j){\frac{\beta\varphi\rho\pi r_j^2 I^*}{\lambda\big(\varphi+\theta(r_j)\big)+\beta\varphi\rho\pi r_j^2 I^*}}-I^*,
\end{equation*}
then we have
\begin{equation*}
\displaystyle  \frac{dF(I^*)}{dI^*}=\sum_{j=1}^m P(r_j){\frac{\beta\varphi\rho\pi r_j^2\big(\lambda\varphi+\lambda\theta(r_j)\big)}{\big(\beta\varphi\rho\pi r_j^2 I^*+\lambda\varphi+\lambda\theta(r_j)\big)^2}}-1,
\end{equation*}
and
\begin{equation*}
\displaystyle  \frac{d^2 F(I^*)}{d{I^*}^2}=\sum_{j=1}^m P(r_j){\frac{-2(\beta\varphi\rho\pi r_j^2)^2\big(\lambda\varphi+\lambda\theta(r_j)\big)}{\big(\beta\varphi\rho\pi r_j^2 I^*+\lambda\varphi+\lambda\theta(r_j)\big)^3}}<0.
\end{equation*}
That is, the continuous function $F(I^*)$ is convex upward in the interval $[0,1]$. In addition, since
\begin{equation*}
\displaystyle  F(0)=0, F(1)=\sum_{j=1}^m P(r_j){\frac{\beta\varphi\rho\pi r_j^2}{\lambda\big(\varphi+\theta(r_j)\big)+\beta\varphi\rho\pi r_j^2}}-1<0,
\end{equation*}
then it follows from the continuity of the function that the necessary and sufficient condition for the existence and uniqueness of the positive solution $0<I^*<1$ to Eq.~(\ref{func8}) should be
\begin{equation*}
\displaystyle  \frac{dF(I^*)}{dI^*}\Bigg|_{I^*=0}=\sum_{j=1}^m P(r_j){\frac{\beta\varphi\rho\pi r_j^2}{\lambda(\varphi+\theta(r_j))}}-1>0.
\end{equation*}
This inequality determines the basic reproduction number $\mathcal{R}_0$ of our model as follows:
\begin{equation}\label{func9}
\displaystyle  \mathcal{R}_0=\frac{\beta\varphi\rho\pi}{\lambda}\sum_{j=1}^m P(r_j){\frac{r_j^2}{\varphi+\theta(r_j)}}.
\end{equation}
In other words, when $\mathcal{R}_0>1$, there exists a unique positive solution $0<I^*<1$ to Eq.~(\ref{func8}) in addition to the trivial solution $I^*=0$; otherwise, the trivial solution $I^*=0$ is the only solution to Eq.~(\ref{func8}). Furthermore, given that $0<I^*<1$, it follows from Eq.~(\ref{func6}) that $0<i^*_{r}<1$ for any $r=r_1,r_2,\dots,r_m$. Therefore, if $\mathcal{R}_0>1$, then the model system (\ref{func4}) has a unique EE point $E^*$ given as
\begin{equation}\label{func10}
E^*=\Big(s^*_{r_1},i^*_{r_1},s^*_{r_2},i^*_{r_2},\dots,s^*_{r_m},i^*_{r_m}\Big),
\end{equation}
where $0<s^*_{r}<1$, $0<i^*_{r}<1$ for any $r=r_1,r_2,\dots,r_m$. We summarize the above statements in the following theorem.

\begin{thm}
Consider the system (\ref{func4}) and the basic reproduction number $\mathcal{R}_0$ defined by Eq.~(\ref{func9}), then there always exists a DFE point $E_0$ as given in Eq.~(\ref{func5}). Moreover, if and only if $\mathcal{R}_0>1$, the system (\ref{func4}) as a unique EE point $E^*$ as given in Eq.~(\ref{func10}). Therefore, the epidemic breaks out when $\mathcal{R}_0>1$; otherwise, the disease dies out eventually.
\end{thm}

\begin{rmk}
In fact, the basic reproduction number $\mathcal{R}_0$ can be interpreted in the epidemiological perspective as follows. Consider a healthy population without any infected seed, then according to the model definition and Eq.~(\ref{func5}), there are only susceptible individuals and vaccinated ones in the steady state, where the stationary number of susceptible individuals in each compartment with effective interaction radius $r=r_1,r_2,\dots,r_m$ is given by $N_S^*(r)=N\frac{\varphi}{\varphi+\theta(r)}$ with the probability distribution $P(r)$. When an infected seed is introduced in the population, the infected individual contacts a susceptible individual with interaction radius $r$ with probability $\pi r^2/D^2$ and transmits the disease to the susceptible at rate $\beta$. Therefore, on average, the infected seed will create $\frac{\beta}{\lambda}\sum_{j=1}^{m}P(r_j)N\frac{\varphi}{\varphi+\theta(r_j)}\frac{\pi r_j^2}{D^2}$ new infections during its entire infectious period $\tau=1/\lambda$. Using the definition of $\rho$ gives $\mathcal{R}_0$.
\end{rmk}

\begin{rmk}
In the extreme case of $\theta(r)=\theta_0$ where a random vaccination scheme is adopted, the basic reproduction number $\mathcal{R}_0$ is given by
\begin{equation*}
\displaystyle  R_0=\frac{\beta\varphi\rho\pi}{\lambda(\varphi+\theta_0)}\sum_{j=1}^m P(r_j)r_j^2,
\end{equation*}
indicating that $\mathcal{R}_0$ is proportional to the transmission rate $\beta$, the population density $\rho$ and the second moment $\langle r^2\rangle=\sum_{j=1}^m P(r_j)r_j^2$ of the radius distribution. In particular, if
$\theta_0=0$, then there is no vaccination and our model reduces to SIS model, where we get $R_0={\beta\rho\pi}\langle r^2\rangle/{\lambda}$, reproducing the result obtained in \cite{Huang16}.
\end{rmk}

\section{Stability analysis} \label{section4}
In this section, we will study the local and global dynamics of DFE point $E_0$ given by (\ref{func5}) and EE point $E^*$ given by (\ref{func10}) of the model system (\ref{func4}). We present all the results on the dynamical behavior of the equilibria in the following theorems.

\subsection{Stability of DFE}

\begin{thm}
Consider the model system (\ref{func4}), the following two conclusions hold.
\begin{itemize}
\item[(1)]If $\mathcal{R}_0<1$, then the DFE point $E_0$ is locally asymptotically stable.
\item[(2)]If $\mathcal{R}_0>1$, then the DFE point $E_0$ is unstable.
\end{itemize}
\end{thm}

\begin{pf}
To determine the local stability of DFE point $E_0$, we consider the Jacobian at the equilibrium $E_0$:
\[ J_{E_0}=\begin{pmatrix}
  A_1 & B_{12} & B_{13} & \dots & B_{1m} \\
  B_{21} & A_2 & B_{23} & \dots & B_{2m} \\
  \vdots & \vdots & \vdots & & \vdots     \\
  B_{m1} & B_{m2} & B_{m3} & \dots & A_m
\end{pmatrix}_{2m\times 2m},               \]
where
\[ A_j=\begin{pmatrix}
  -\Big(\theta(r_j)+\varphi\Big) & \lambda-\varphi-\frac{\varphi}{\varphi+\theta(r_j)}\beta\rho\pi r_j^2 P(r_j) \\
  0 & -\lambda+\frac{\varphi}{\varphi+\theta(r_j)}\beta\rho\pi r_j^2 P(r_j)
\end{pmatrix}, \quad j=1,2,\dots,m,   \]
and
\[ B_{ij}=\begin{pmatrix}
  0 & -\frac{\varphi}{\varphi+\theta(r_i)}\beta\rho\pi r_i^2 P(r_j) \\
  0 & \frac{\varphi}{\varphi+\theta(r_i)}\beta\rho\pi r_i^2 P(r_j)
\end{pmatrix}, \quad i,j=1,2,\dots,m.  \]
Then the characteristic equation of the Jacobian matrix $J_{E_0}$ is
\begin{equation*}
\bigg[\prod_{j=1}^m \Big(x+\varphi+\theta(r_j)\Big)\bigg](x+\lambda)^{m-1}\Big[x+\lambda-\beta\varphi\rho\pi{\sum_{j=1}^m P(r_j){\frac{r_j^2}{\varphi+\theta(r_j)}}}\Big]=0,
\end{equation*}
where the variable $x$ denotes eigenvalues of the matrix $J_{E_0}$. Obviously, when $\mathcal{R}_0<1$, all the eigenvalues are negative and the DFE point $E_0$ is locally asymptotically stable. Otherwise, when $\mathcal{R}_0>1$, there is a positive eigenvalue, suggesting that the DFE point $E_0$ is unstable. This concludes the proof. \qed

\end{pf}

Furthermore, in the following theorem we provide a sufficient condition to guarantee the global asymptotically stability of DFE.

\begin{thm}
If $\mathcal{R}_0<1$ and $\lambda\leq\varphi$, then the DFE point $E_0$ of model (\ref{func4}) is globally asymptotically stable.
\end{thm}

\begin{pf}
Suppose $\lambda\leq\varphi$, then it follows from the first equation of model (\ref{func4}) that for any $r=r_1,r_2,\dots,r_m$, we have
\begin{equation*}
\begin{split}
\frac{ds(r,t)}{dt} &= \lambda i(r,t)-s(r,t)\beta\rho\pi r^2 I(t)+\varphi \Big[1-s(r,t)-i(r,t)\Big]-s(r,t)\theta(r) \\
                   &= \varphi-\varphi s(r,t)-\varphi i(r,t)-s(r,t)\theta(r)+\lambda i(r,t)-s(r,t)\beta\rho\pi r^2 I(t) \\
                   &\leq \varphi-\Big(\varphi+\theta(r)\Big)s(r,t)+(\lambda-\varphi)i(r,t) \\
                   &\leq \varphi-\Big(\varphi+\theta(r)\Big)s(r,t). \\
\end{split}
\end{equation*}
Hence, $s(r,t)\leq \frac{\varphi}{\varphi+\theta(r)}$ for any $r=r_1,r_2,\dots,r_m$ and $t\geq 0$. Define the Lyapunov function as
\begin{equation*}
L(t)=I(t)=\sum_{j=1}^{m} P(r_j)i(r_j,t).
\end{equation*}
Then, the derivative of $L(t)$ along the system (\ref{func4}) is given as
\begin{equation*}
\begin{split}
L'(t)&=\frac{d}{dt}L(t) =\sum_{j=1}^{m} P(r_j)\frac{d}{dt}i(r_j,t)= \sum_{j=1}^{m} P(r_j)\Big[-\lambda i(r_j,t)+s(r_j,t)\beta\rho\pi r_j^2 I(t)\Big] \\
      &= -\lambda \sum_{j=1}^{m} P(r_j)i(r_j,t)+\sum_{j=1}^{m} P(r_j)s(r_j,t)\beta\rho\pi r_j^2 I(t) \\
      &\leq -\lambda I(t)+\sum_{j=1}^{m} P(r_j)\frac{\varphi}{\varphi+\theta(r_j)} \beta\rho\pi r_j^2 I(t) \\
      &= I(t)\bigg[\beta\varphi\rho\pi \sum_{j=1}^{m} P(r_j)\frac{r_j^2}{\varphi+\theta(r_j)}-\lambda\bigg]. \\
\end{split}
\end{equation*}
Note that $\mathcal{R}_0<1$ is equivalent to $\beta\rho\varphi\pi {\sum_{j=1}^{m} P(r_j){\frac{r_j^2}{\varphi+\theta(r_j)}}}-\lambda<0$. Therefore, if $\lambda\leq\varphi$ and $\mathcal{R}_0<1$, then $L'(t) \leq 0$. Furthermore, $L'(t)=0$ if and only if $i(r,t)=0$ for any $r=r_1,r_2,\dots,r_m$. By LaSalle's invariance principle \cite{LaSalle76}, we conclude that if $\mathcal{R}_0<1$ and $\lambda\leq\varphi$, then the DFE point $E_0$ is globally asymptotically stable. This completes the proof. \qed
\end{pf}

\subsection{Stability of EE}

\begin{thm}
If $\mathcal{R}_0>1$, then the system (\ref{func4}) admits a unique EE point $E^*$ defined by Eq.~(\ref{func10}) which is locally asymptotically stable in $\Gamma$.
\end{thm}

\begin{pf}
Define $\hat{s}(r,t)=s(r,t)-s_r^*$ and $\hat{i}(r,t)=i(r,t)-i_r^*$ for each $r=r_1,r_2,\dots,r_m$. We consider the following linearized dynamics of system (\ref{func4}) at $E^*$.
\begin{equation}\label{func12a}
\begin{array}{ll}
\displaystyle  \frac{d\hat{s}(r,t)}{dt}=-\Big[\beta\rho\pi r^2 I^*+\theta(r)+\varphi\Big]\hat{s}(r,t)+(\lambda-\varphi)\hat{i}(r,t)-s_r^*\beta\rho\pi r^2 \hat{I}(t),  \\
\displaystyle  \frac{d\hat{i}(r,t)}{dt}=\hat{s}(r,t)\beta\rho\pi r^2 I^*+s_r^*\beta\rho\pi r^2 \hat{I}(t)-\lambda \hat{i}(r,t), \quad r=r_1,r_2,\dots,r_m, \\
\end{array}
\end{equation}
where $\hat{I}(t)=\sum\limits_{j=1}^m P(r_j)\hat{i}(r_j,t)$ and $I^*=\sum\limits_{j=1}^m P(r_j)i_{r_j}^*$.

Suppose $\xi$ is an arbitrary eigenvalue of the coefficient matrix of the linearized system \eqref{func12a}. Then the proof can be done if we verify that $\xi$ has negative real part, i.e., $Re{(\xi)}<0$. Looking for exponential solutions of the linear equations \eqref{func12a}, we set $\hat{s}(r,t)=\hat{s}_0(r)e^{\xi t}$ and $\hat{i}(r,t)=\hat{i}_0(r)e^{\xi t}$ for any $r=r_1,r_2,\dots,r_m$. Substituting in the linearized system and canceling $e^{\xi t}$, we obtain
\begin{equation}\label{func12b}
\begin{array}{ll}
\displaystyle \xi \hat{s}_0(r)=-[\beta\rho\pi r^2 I^*+\theta(r)+\varphi]\hat{s}_0(r)+(\lambda-\varphi)\hat{i}_0(r)-s_r^*\beta\rho\pi r^2 \sum_{j=1}^m P(r_j)\hat{i}_0(r_j), \\
\displaystyle \xi \hat{i}_0(r)=\hat{s}_0(r)\beta\rho\pi r^2 I^*+s_r^*\beta\rho\pi r^2 \sum_{j=1}^m P(r_j)\hat{i}_0(r_j)-\lambda \hat{i}_0(r). \\
\end{array}
\end{equation}
Thus
\begin{equation}\label{func13}
\begin{array}{ll}
\displaystyle \hat{s}_0(r)=\frac{(\lambda-\varphi)\hat{i}_0(r)-s_r^*\beta\rho\pi r^2 \sum\limits_{j=1}^m P(r_j)\hat{i}_0(r_j)}{\xi+\beta\rho\pi r^2 I^*+\theta(r)+\varphi}, \\
\displaystyle \hat{i}_0(r)=\frac{\hat{s}_0(r)\beta\rho\pi r^2 I^*+s_r^*\beta\rho\pi r^2 \sum\limits_{j=1}^m P(r_j)\hat{i}_0(r_j)}{\xi+\lambda}. \\
\end{array}
\end{equation}
By denoting $Y(\hat{i}_0)=\sum\limits_{j=1}^m P(r_j)\hat{i}_0(r_j)=\hat{I}(t)e^{-\xi t}$, we get
\begin{equation}\label{func14}
\hat{i}_0(r)=\frac{s_r^*\beta\rho\pi r^2[\xi+\theta(r)+\varphi]}{(\xi+\lambda)[\xi+\theta(r)+\varphi]+(\xi+\varphi)\beta\rho\pi r^2 I^*}Y(\hat{i}_0).
\end{equation}
Multiplying Eq.~\eqref{func14} by $P(r)e^{\xi t}$ and summarizing over all $r=r_1,r_2,\dots,r_m$ gives rise to
\begin{equation}\label{func15}
\hat{I}(t)=\sum_{j=1}^m P(r_j)\frac{s_{r_j}^*\beta\rho\pi r_j^2 [\xi+\theta(r_j)+\varphi]}{(\xi+\lambda)[\xi+\theta(r_j)+\varphi]+(\xi+\varphi)\beta\rho\pi r_j^2 I^*}\hat{I}(t).
\end{equation}
If $\hat{I}(t)=0$, then $Y(\hat{i}_0)=0$ and hence $\hat{i}_0(r)=0$ by Eq.~\eqref{func14}. In this case, we obtain from the first equation of \eqref{func12b} that
\begin{equation*}
Re(\xi)=\xi=-[\beta\rho\pi r^2 I^*+\theta(r)+\varphi]<0.
\end{equation*}
If $\hat{I}(t)\neq 0$, then according to Eq.~\eqref{func15} we have
\begin{equation}\label{func16}
\sum_{j=1}^m P(r_j)\frac{s_{r_j}^*\beta\rho\pi r_j^2 }{\lambda+\xi+\frac{\xi+\varphi}{\xi+\theta(r_j)+\varphi}\beta\rho\pi r_j^2 I^*}=1.
\end{equation}
Letting the right-hand side of the second equation of system \eqref{func4} be zero, we obtain a stationary solution
\begin{equation*}
\lambda i_r^*-s_r^*\beta\rho\pi r^2 I^*=0, \quad r=r_1,r_2,\dots,r_m.
\end{equation*}
By Eq.~\eqref{func7} we obtain
\begin{equation*}
\lambda I^*-I^*\beta\rho\pi \sum_{j=1}^m P(r_j)s_{r_j}^* r_j^2=0.
\end{equation*}
Note that $I^*>0$ as long as $\mathcal{R}_0>1$, therefore
\begin{equation}\label{func17}
\beta\rho\pi \sum_{j=1}^m P(r_j)s_{r_j}^* r_j^2=\lambda.
\end{equation}
Next, we verify that $Re(\xi)<0$ by the way of contradiction.

(i) if $\xi=0$, then it follows from Eq.~\eqref{func16} that
\begin{equation}\label{func18}
\sum_{j=1}^m P(r_j)\frac{s_{r_j}^*\beta\rho\pi r_j^2}{\lambda+\frac{\varphi}{\theta(r_j)+\varphi}\beta\rho\pi r_j^2 I^*}=1.
\end{equation}
Since $I^*>0$, Eq.~\eqref{func18} indicates that
\begin{equation*}
1=\sum_{j=1}^m P(r_j)\frac{s_{r_j}^*\beta\rho\pi r_j^2}{\lambda+\frac{\varphi}{\theta(r_j)+\varphi}\beta\rho\pi r_j^2 I^*}<\sum_{j=1}^m P(r_j)\frac{s_{r_j}^*\beta\rho\pi r_j^2}{\lambda}=1,
\end{equation*}
which is a contradiction. Hence, $\xi\neq 0$.

(ii) if $Re(\xi)>0$, then $Re(\xi+\frac{\xi+\varphi}{\xi+\theta(r_j)+\varphi}\beta\rho\pi r_j^2 I^*)>0$ $(j=1,2,\dots,m)$. Let
\begin{equation*}
\xi+\frac{\xi+\varphi}{\xi+\theta(r_j)+\varphi}\beta\rho\pi r_j^2 I^*=a+b\mathbf{i} \quad (a>0),
\end{equation*}
where $\mathbf{i}$ is the imaginary unit. Consequently, we have
\begin{equation*}
\frac{s_{r_j}^*\beta\rho\pi r_j^2}{\lambda+\xi+\frac{\xi+\varphi}{\xi+\theta(r_j)+\varphi}\beta\rho\pi r_j^2 I^*}=\frac{s_{r_j}^*\beta\rho\pi r_j^2}{\lambda+a+b\mathbf{i}}=\frac{s_{r_j}^*\beta\rho\pi r_j^2(\lambda+a)-(s_{r_j}^*\beta\rho\pi r_j^2 b)\mathbf{i}}{(\lambda+a)^2+b^2}.
\end{equation*}
Therefore,
\begin{equation*}
Re\bigg(\frac{s_{r_j}^*\beta\rho\pi r_j^2}{\lambda+\xi+\frac{\xi+\varphi}{\xi+\theta(r_j)+\varphi}\beta\rho\pi r_j^2 I^*}\bigg)=\frac{s_{r_j}^*\beta\rho\pi r_j^2(\lambda+a)}{(\lambda+a)^2+b^2}\leq \frac{s_{r_j}^*\beta\rho\pi r_j^2}{\lambda+a}<\frac{s_{r_j}^*\beta\rho\pi r_j^2}{\lambda}.
\end{equation*}
This implies
\begin{equation*}
\begin{split}
&Re\bigg(\sum_{j=1}^m P(r_j)\frac{s_{r_j}^*\beta\rho\pi r_j^2}{\lambda+\xi+\frac{\xi+\varphi}{\xi+\theta(r_j)+\varphi}\beta\rho\pi r_j^2 I^*}\bigg) \\
&<\sum_{j=1}^m P(r_j)\frac{s_{r_j}^*\beta\rho\pi r_j^2}{\lambda}=1, \\
\end{split}
\end{equation*}
which gives a contradiction to Eq.~\eqref{func16}. Hence, $Re(\xi)\leq0$.

(iii) If $Re(\xi)=0$, then the imaginary part $Im(\xi)\neq 0$ according to (i). In this case, we also find that $Re(\xi+\frac{\xi+\varphi}{\xi+\theta(r_j)+\varphi}\beta\rho\pi r_j^2 I^*)>0$ $(j=1,2,\dots,m)$. This results in a contradiction with Eq.~\eqref{func16} again in an analogous fashion. Thus, $Re(\xi)<0$.

Based on the above discussion, we conclude that if $\mathcal{R}_0>1$, then all the eigenvalues have negative real part, thus the EE point $E^*$ of model \eqref{func4} is locally asymptotically stable. This completes the proof. \qed

\end{pf}

\section{Persistence of the disease}\label{section5}
As pointed out in the previous section, there is an EE as long as $\mathcal{R}_0>1$. In this section we additionally present the following theorem on the persistence of the disease in the case of $\mathcal{R}_0>1$.
\begin{thm}
If $\mathcal{R}_0>1$, then the system \eqref{func4} is persistent, that is, there exists $\varepsilon>0$ such that
\begin{equation*}
\liminf_{t\to\infty} I(t)=\liminf_{t\to\infty} \sum_{j=1}^m P(r_j)i(r_j,t)>\varepsilon.
\end{equation*}
\end{thm}

\begin{pf}
We will use the conclusion given by Thieme (see theorem 4.6 in \cite{Thieme93}) to prove the above proposition. Starting with the positively invariant set $\Gamma$ given by \eqref{func4b}, we define two sets
\begin{equation*}
\begin{split}
&\Gamma^*=\bigg\{\Big(s(r_1,t),i(r_1,t),\dots,s(r_m,t),i(r_m,t)\Big)\in \Gamma\bigg{|}\sum_{j=1}^m P(r_j)i(r_j,t)>0\bigg\}, \\
&\partial \Gamma^*=\Gamma\backslash \Gamma^*=\bigg\{\Big(s(r_1,t),i(r_1,t),\dots,s(r_m,t),i(r_m,t)\Big)\in \Gamma\bigg{|}\sum_{j=1}^m P(r_j)i(r_j,t)=0\bigg\}.
\end{split}
\end{equation*}

According to the proof of Lemma \ref{lem1}, if $s(r,0)\geq 0, v(r,0)\geq 0, I(0)=\sum\limits_{j=1}^m P(r_j)i(r_j,0)>0$, then $s(r,t)\geq 0, v(r,t)\geq 0, I(t)=\sum\limits_{j=1}^m P(r_j)i(r_j,t)>0$ for any $t>0$. Note that
\begin{equation*}
I'(t)=\sum_{j=1}^m P(r_j)\Big(s(r_j,t)\beta\rho\pi r_j^2 I(t)-\lambda i(r_j,t)\Big) \geq -\lambda I(t),
\end{equation*}
and $I(0)=\sum\limits_{j=1}^m P(r_j)i(r_j,0)>0$, it follows from the comparison theorem \cite{Lakshmikantham15} that
\begin{equation*}
I(t)=\sum_{j=1}^m P(r_j)i(r_j,t) \geq \sum_{j=1}^m P(r_j)i(r_j,0)e^{-\lambda t}>0.
\end{equation*}
Hence, the set $\Gamma^*$ is also positively invariant. Moreover, there exists a compact set $B$ in which all solutions to \eqref{func4} starting from $\Gamma$ will enter and remain permanently. It can be easily validated that the set $B$ satisfies the compactness conditions ($C_{4.2}$) proposed by Thieme \cite{Thieme93}. Denote
\begin{equation*}
W_\partial=\Big\{\Big(s(r,0),i(r,0)\Big)\bigg{|} \Big(s(r,t),i(r,t)\Big) \in \partial \Gamma^*, r=r_1,r_2,\dots,r_m,t\geq 0\Big\},
\end{equation*}
and
\begin{equation*}
\Omega_\Gamma=\bigcup\limits_{(s(r,0),i(r,0)) \in \Gamma} \omega\Big(s(r,0),i(r,0)\Big),
\end{equation*}
where $\omega \Big(s(r,0),i(r,0)\Big)$ is the $\omega$-limit set of the solutions of system \eqref{func4} initiated from $(s(r,0),i(r,0))$. Confining the system \eqref{func4} to $W_\partial$ yields

\begin{equation}\label{func12}
\begin{array}{ll}
\displaystyle  \frac{ds(r,t)}{dt}=\lambda i(r,t)+\varphi [1-s(r,t)-i(r,t)]-s(r,t)\theta(r), \\
\displaystyle  \frac{di(r,t)}{dt}=-\lambda i(r,t).
\end{array}
\end{equation}
It is easy to manifest that the system \eqref{func12} has a unique equilibrium $E_0$ as denoted by Eq.~\eqref{func5}, which is globally asymptotically stable. Thus, $\Omega_\Gamma=\{E_0\}$. In addition, $E_0$ is an acyclic isolated covering of $\Omega_\Gamma$ since there is no solution in $W_\partial$ that links $E_0$ to itself. Next, we will prove that $\{E_0\}$ is a weak repeller for $\Gamma^*$, namely, any solution $(s(r,t),i(r,t))$ with initial value in $\Gamma^*$ satisfies
\begin{equation*}
\limsup_{t\to\infty} dist\Big(\big(s(r,t),i(r,t)\big),\{E_0\}\Big)>0.
\end{equation*}
Here, the distance $dist(x,Y)$ of a point $x\in X$ from a subset $Y$ of $X$ is defined by
\begin{equation*}
dist(x,Y)=\inf\limits_{y\in Y}d(x,y),
\end{equation*}
where $d$ is a metric of the metric space $X$.
According to the proof of lemma 3.5 by Leenheer and Smith \cite{Smith03}, we only need to verify $M^s(E_0)\cap \Gamma^*=\emptyset$, where $M^s(E_0)$ is the stable manifold of $E_0$. We prove it by the way of contradiction. Suppose $M^s(E_0)\cap \Gamma^*\neq\emptyset$, then there exists a solution $(s(r,t),i(r,t))\in \Gamma^*$ such that
\begin{equation}\label{func13}
s(r,t)\to \frac{\varphi}{\varphi+\theta(r)}, \quad i(r,t)\to 0 \quad \text{as} \quad t\to\infty.
\end{equation}
It is clear form \eqref{func13} that for any given $\eta>0$, there exists $T_0>0$ such that
\begin{equation*}
\frac{\varphi}{\varphi+\theta(r)}-\eta<s(r,t)<\frac{\varphi}{\varphi+\theta(r)}+\eta, \quad 0\leq i(r,t)<\eta \quad \text{as} \quad t\geq T_0.
\end{equation*}
Given that $\mathcal{R}_0=\frac{\beta\rho\pi}{\lambda}\sum\limits_{j=1}^m P(r_j)r_j^2{\frac{\varphi}{\varphi+\theta(r_j)}}>1$, there exists a positive constant $\eta>0$ such that
\begin{equation*}
\frac{\beta\rho\pi}{\lambda}\sum\limits_{j=1}^m P(r_j)r_j^2\Big({\frac{\varphi}{\varphi+\theta(r_j)}}-\eta\Big)>1.
\end{equation*}
Let
\begin{equation*}
C=\beta\rho\pi\sum\limits_{j=1}^m P(r_j)r_j^2\Big({\frac{\varphi}{\varphi+\theta(r_j)}}-\eta\Big)-\lambda,
\end{equation*}
thus $C>0$ as long as $\mathcal{R}_0>1$. Define the following Lyapunov function
\begin{equation*}
V(t)=I(t)=\sum_{j=1}^m P(r_j)i(r_j,t),
\end{equation*}
then the derivative of $V$ along the solution $(s(r,t),i(r,t))$ of system \eqref{func4} is given by
\begin{equation*}
\begin{split}
\frac{dV(t)}{dt}\Big{|}_{\eqref{func4}} &= \sum_{j=1}^m P(r_j)\big[-\lambda i(r_j,t)+s(r_j,t)\beta\rho\pi r_j^2\sum_{\ell=1}^m P(r_\ell)i(r_\ell,t) \big] \\
&= \sum_{j=1}^m \Big[\beta\rho\pi\sum\limits_{j=1}^m P(r_j)r_j^2 s(r_j,t)-\lambda\Big]P(r_j)i(r_j,t).
\end{split}
\end{equation*}
It follows that for all $t\geq T_0$,
\begin{equation*}
\begin{split}
\frac{dV(t)}{dt}\Big{|}_{\eqref{func4}} \geq \sum_{j=1}^m \Big[\beta\rho\pi\sum\limits_{j=1}^m P(r_j)r_j^2\Big({\frac{\varphi}{\varphi+\theta(r_j)}}-\eta\Big)-\lambda\Big]P(r_j)i(r_j,t).
\end{split}
\end{equation*}
That is, $\frac{dV(t)}{dt}\Big{|}_{\eqref{func4}}\geq CV(t)$ for all $t\geq T_0$, which implies $\lim\limits_{t\to\infty} V(t)=\infty$. This is a contradiction to the boundedness of $V(t)$. Therefore, $\{E_0\}$ is a weak repeller for $\Gamma^*$. Based on the result by Thieme (theorem 4.6 in \cite{Thieme93}), we conclude that if $\mathcal{R}_0>1$ then the system \eqref{func4} is persistent. \qed

\end{pf}

\section{Simulation results and discussion} \label{section6}

In this section, we provide extensive stochastic simulations to support the theoretical results of our model. In the end we also display some simulation results on the topological structure of the underlying contact network.

\subsection{Spreading dynamics of the epidemic model}

In our simulations, the length of the side of the square space is set to be $D=30$ and the velocity $v$ of each individual is fixed to be $v=0.1$. We start the simulation with $1\%$ of individuals being infected seeds while the others all susceptible. That is, at $t=0$ the number of infected, susceptible and vaccinated individuals is $N_I(0)=9$, $N_S(0)=891$ and $N_V(0)=0$, respectively. At the beginning, all the individuals are randomly distributed within the planar space, where at each time step the individuals move depending on the random jump probability $p_{\text jump}$. As long as an infected individual runs into an effective interaction radius of a susceptible one, an infection takes place at the transmission rate $\beta$. The simulation is ended as the number (or fraction) of each class of individuals reaches a relatively steady level (with negligible fluctuations). To investigate the impacts of heterogeneity in the interaction radius among individuals, we consider different distributions of interaction radius in the model, including the Poisson, exponential, power-law and in some cases the Kronecker delta.

\begin{figure}[htp]
\centering
\includegraphics[scale=0.32]{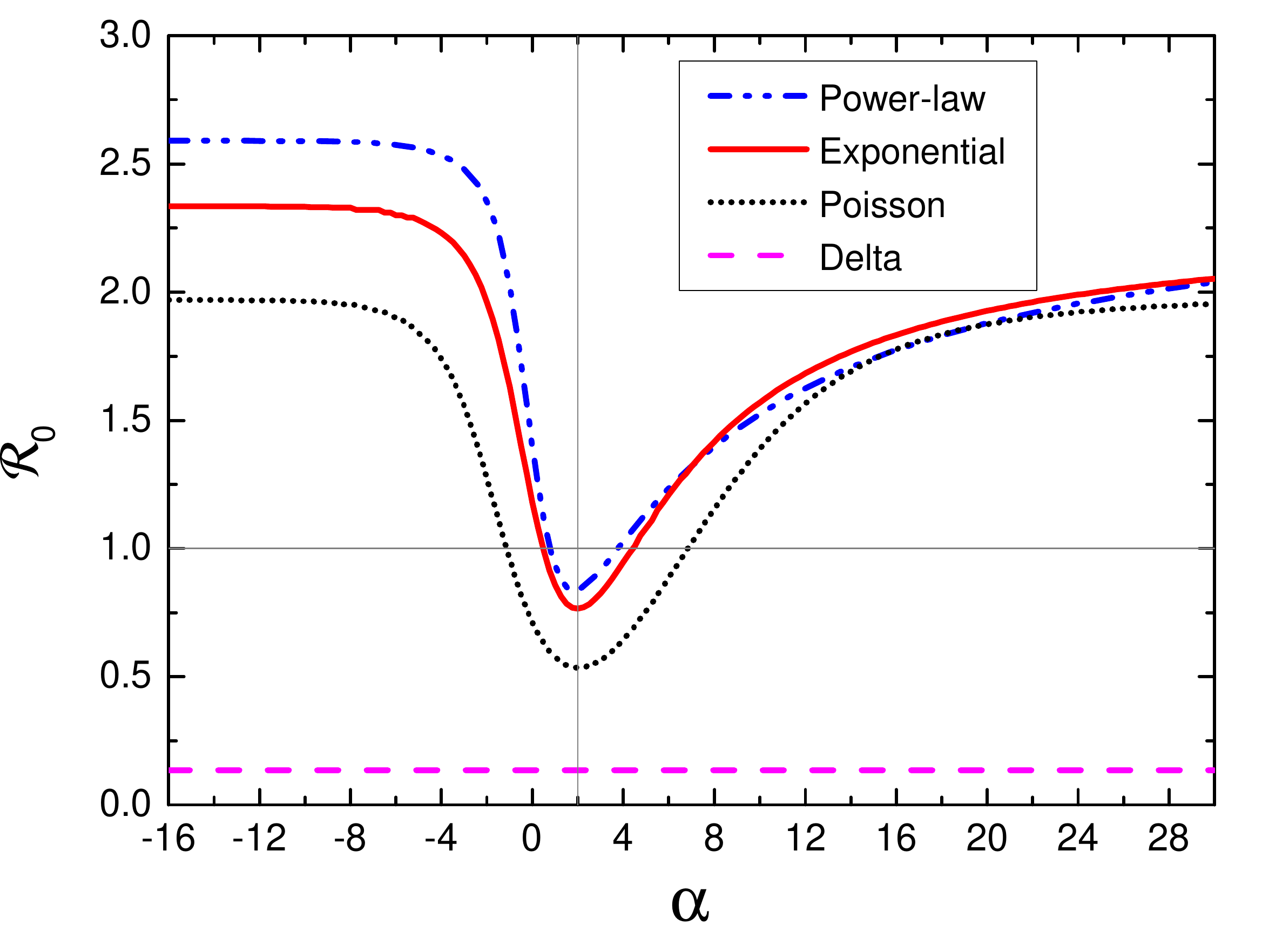}
\caption{(Color online) The basic reproduction number $\mathcal{R}_0$ as a function of the tunable parameter $\alpha$. The values are obtained by Eq.~(\ref{func9}). Four different distributions $P(r)$ of interaction radius have been adopted: power-law (blue, dash-dot-dotted lines), exponential (red, solid lines), Poisson (black, dotted lines) and Kronecker Delta (pink, dashed lines). The former three distributions take the values of radius $r\in\{1,2,3,4,5,6,7,8,9,10\}$ with the same average $\langle r\rangle=3$ whereas the last one takes $r\equiv3$ (namely, $P(r)=\delta_{r3}$). The grey thin lines are presented as a clear guide: the vertical line corresponding to $\alpha=2$ and the horizontal line corresponding to $\mathcal{R}_0=1$. Other parameters are $N=900, D=30, \rho=1, v=0.1, \beta=0.005, \lambda=0.05, \varphi=0.005, \theta_0=0.1, m=10$.}\label{fig4}
\end{figure}

Figure~\ref{fig4} exhibits a comparison of the value of the basic reproduction number $\mathcal{R}_0$ as a function of the vaccination-strength-related parameter $\alpha$ between four different distributions $P(r)$ of interaction radius of individuals. In the case of the Kronecker Delta distribution $P(r)=\delta_{r3}$ (where $\delta_{r3}=1$ if $r=3$ and $\delta_{r3}=0$ otherwise), all the individuals have an identical interaction radius, thus the vaccination rate $\theta(r)$ turns to be $\theta(r)=\theta(3)=\theta_0$ which is a constant. This indicates the value of $\mathcal{R}_0$ is a constant, in particular by Eq.~(\ref{func9}), $\mathcal{R}_0=3^2\beta\varphi\rho\pi/[\lambda(\varphi+\theta_0)]\simeq0.135$. In the case of Poisson, exponential and power-law distributions, the value of $\mathcal{R}_0$ remains almost unchanged for $\alpha<-5$. This is because the smaller the value of parameter $\alpha$, the larger the vaccination probability for individuals with small radius. Therefore, as $\alpha$ is small enough, the individuals to be vaccinated are only those with the smallest radius since their vaccination probability is dominantly large. When $-5<\alpha<2$, the value of $\mathcal{R}_0$ decreases with $\alpha$ drastically and in general the value of $\mathcal{R}_0$ in the power-law radius distribution is larger than the value of $\mathcal{R}_0$ in the exponential radius distribution, which is in turn greater than that in the Poisson radius distribution. When $\alpha>2$, the value of $\mathcal{R}_0$ first increases relatively fast and then grows gradually slowly with large $\alpha$. Again, the values of $\mathcal{R}_0$ in power-law and exponential radius distributions are greater than that in the Poisson and Delta radius distributions. This result suggests that it is easier for the disease to break out in the population with more heterogeneous distribution of interaction radius due to a larger $\mathcal{R}_0$. It is interesting to notice that for all of the power-law, exponential and Poisson distributions of interaction radius, the value of $\mathcal{R}_0$ reaches a minimum at $\alpha=2$, as illustrated by the grey vertical line in Fig.~\ref{fig4}. This may motivate an optimal vaccination intervention for disease prevention irrespective of the distribution of individuals' effective interaction radius.

\begin{figure}[htp]
\centering
\includegraphics[scale=0.5]{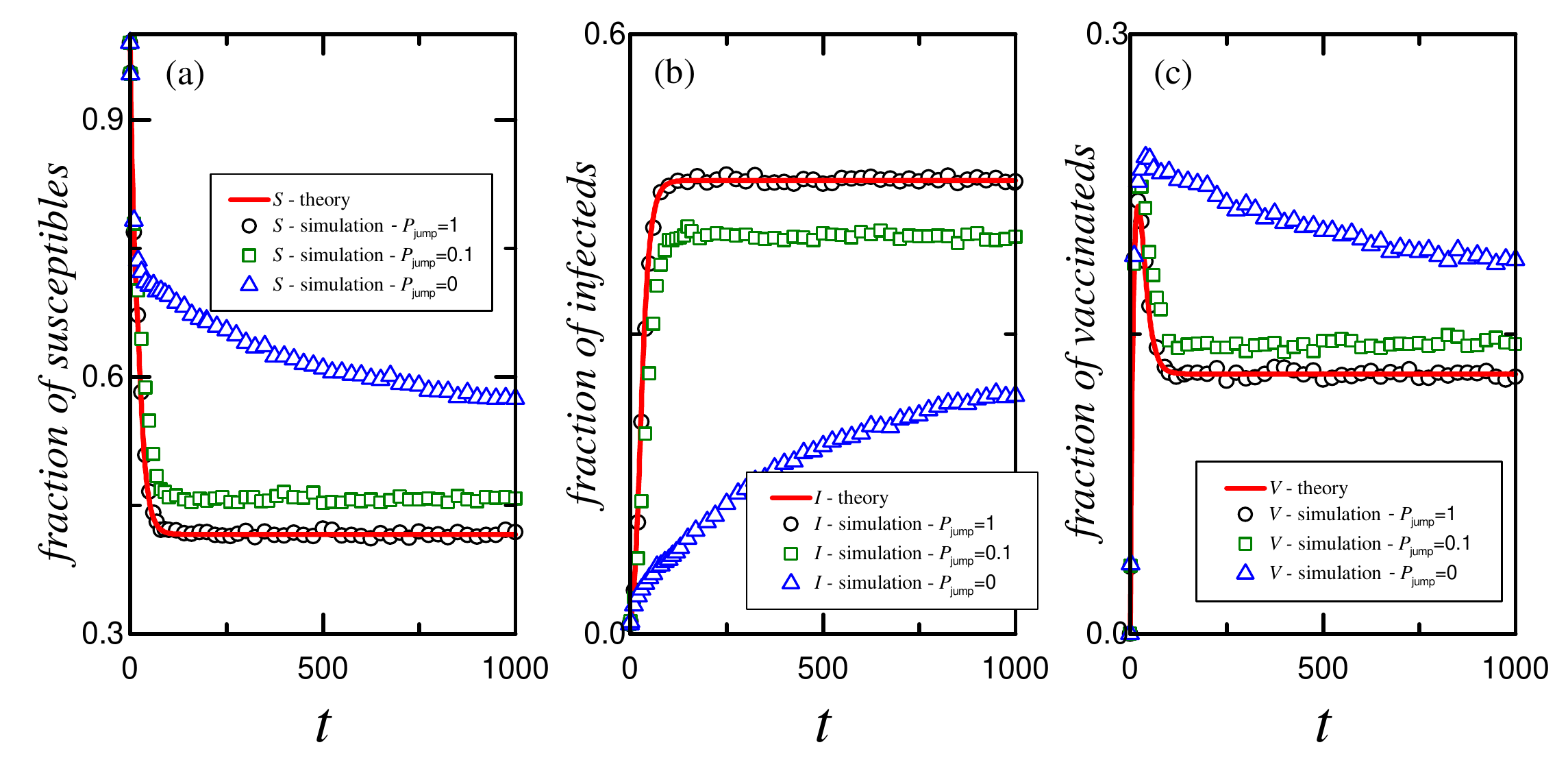}
\caption{(Color online) Time series of the fractions of susceptible (a), infected (b) and vaccinated (c) individuals in the population for different values of $p_{\text jump}$. Solid (red) lines correspond to the results obtained from model (\ref{func2}) based on the HM hypothesis. The empty (black) circles, (olive) squares and (blue) triangles represent the stochastic simulation results for $p_{\text jump}=1, 0.1$ and $0$, respectively. Each of the simulation point has been averaged over 50 independent realizations. Here, we take a simple interaction radius distribution with $m=3$, $r_1=0.5, r_2=1, r_3=1.5$ and $P(r_1)=0.3, P(r_2)=0.4$ and $P(r_3)=0.3$. Other parameters are $N=900, D=30, \rho=1, v=0.1, \beta=0.08, \lambda=0.1, \varphi=0.2, \theta_0=0.1, \alpha=0.5$ and $\mathcal{R}_0=2.4304$.}\label{fig5}
\end{figure}

Figure \ref{fig5} depicts the time series of the density of each class of individuals in the population for different values of $p_{\text jump}$ given other parameters. It is clear from Fig.~\ref{fig5}(b) that the density of infected individuals grows with $p_{\text jump}$, while it is shown from Figs.~\ref{fig5}(a, c) that both the densities of susceptible individuals and vaccinated ones decrease with $p_{\text jump}$. This means that for larger $p_{\text jump}$ the infection is more severe \cite{Frasca06,Buscarino10,Xia10,Huang16}. Moreover, the simulation results in the case with $p_{\text jump}=1$ are in perfect agreement with the theoretical predictions based on the HM assumption, which confirms the statement ahead of model (\ref{func1}). As done in Ref.~\cite{Huang16}, in what follows we only investigate the case of $p_{\text jump}=1$ since in other cases our model (\ref{func1}) based on HM assumption deviates obviously from the simulation results, as demonstrated in Fig.~\ref{fig5}.

\begin{figure}[htp]
\centering
\includegraphics[scale=0.3]{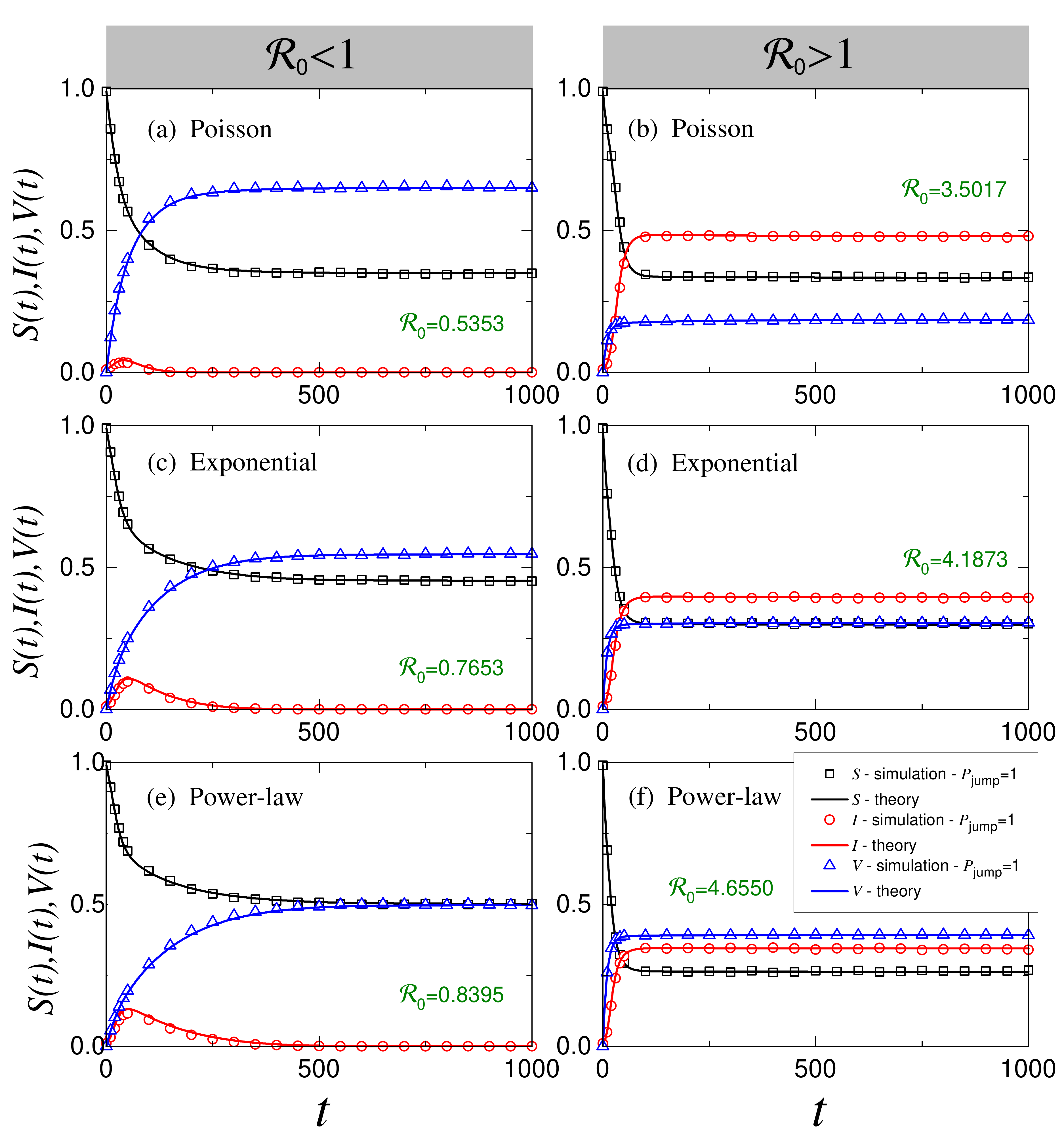}
\caption{(Color online) Time series of the fractions of susceptible ($S(t)$, marked in black), infected ($I(t)$, marked in red) and vaccinated ($V(t)$, marked in blue) individuals in the population under the conditions that $\mathcal{R}_0<1$ with $\lambda=0.09, \alpha=2$ (panels (a), (c), (e)) and that $\mathcal{R}_0>1$ with $\lambda=0.05, \alpha=-8$ (panels (b), (d), (f)). Three distributions of interaction radius have been included: (a), (b) for Poisson; (c), (d) for exponential and (e), (f) for power-law. All of these distributions take the radius sequence $r\in\{1,2,3,4,5,6,7,8,9,10\}$ with the same average $\langle r\rangle=3$. Solid lines stand for theoretical results by system (\ref{func2}) and (\ref{func7}) and each symbol (namely, square, circle and triangle) corresponds to the stochastic simulation result that is obtained by averaging over 100 realizations. Other parameters are $N=900, D=30, \rho=1, v=0.1, \beta=0.005, \varphi=0.005, \theta_0=0.1, m=10$. Note that the value of $\mathcal{R}_0$ has been denoted in green in each panel.}\label{fig6}
\end{figure}

\begin{figure}[htp]
\centering
\includegraphics[scale=0.35]{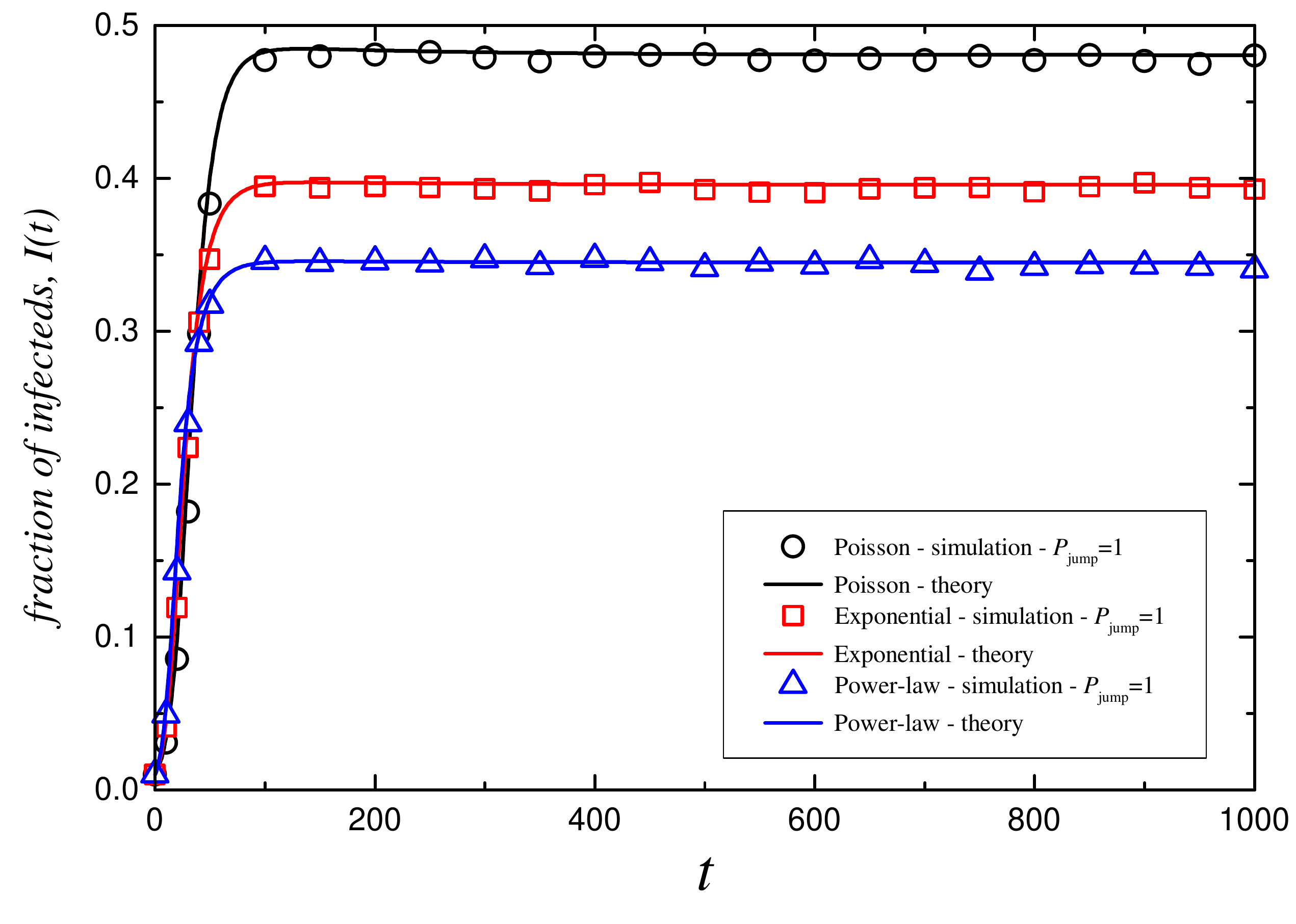}
\caption{(Color online) Comparison of the time evolution of the fraction of infected individuals in the population between different distributions of interaction radius. All the data and parameters are as in Figs.~\ref{fig6}(b,d,f). Solid lines are the theoretical results and symbols are stochastic simulation results. The Poisson, exponential and power-law distributions are colored in black, red and blue, respectively.}\label{fig7}
\end{figure}

Figure \ref{fig6} plots the temporal evolution of the density of susceptible, infected and vaccinated individuals in the population for different distributions of effective interaction radius. Both the results under the condition of $\mathcal{R}_0<1$ and in the case of $\mathcal{R}_0>1$ have been calculated. When $\mathcal{R}_0<1$, the fraction of infected population eventually decays to zero irrespective of the radius distribution. Otherwise, when $\mathcal{R}_0>1$, as shown in Figs.~\ref{fig6}(b,d,f) the density of infected individuals raises swiftly in the early stage and finally enters a stationary state. It is demonstrated that all the simulation results in the case of $p_{\text jump}=1$ agree well with the theoretical predictions by our model based on the HM assumption. To further explore the effects of the heterogeneity in the interaction radius on the epidemic spread, we give a clear comparison of the density of infected individuals between the Poisson, exponential and power-law radius distributions in Fig.~\ref{fig7}, where the data is extracted from Figs.~\ref{fig6}(b,d,f). It is shown from Fig.~\ref{fig7} that the final epidemic prevalence in the power-law interaction radius distribution is at the lowest level although the basic reproduction number $\mathcal{R}_0=4.655$ in the power-law case is the largest among the three distributions. On the contrary, the final epidemic prevalence (i.e. the final fraction of infected individuals) in the case of Poisson radius distribution is the highest, albeit with the smallest basic reproduction number $\mathcal{R}_0=3.5017$. This interesting result reveals that the disease will break out readily in the population with a heterogeneous interaction radius distribution; however, resulting in a relatively small epidemic prevalence.

\begin{figure}[htp]
\centering
\includegraphics[scale=0.3]{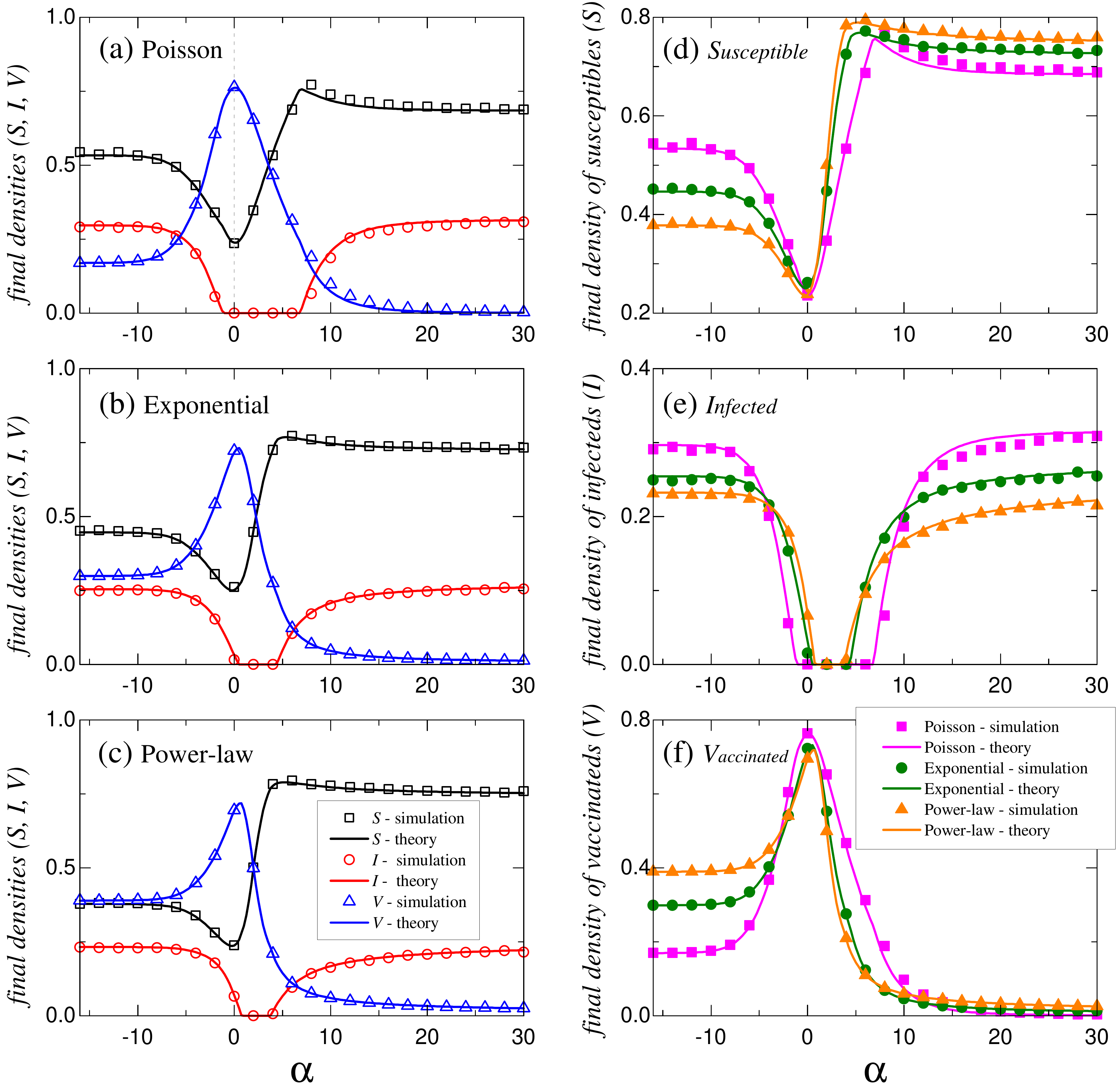}
\caption{(Color online) The final densities of susceptible ($S$), infected ($I$) and vaccinated ($V$) individuals in the steady state as a function of $\alpha$. The left column panels (a, b, c) are sorted by different interaction radius distributions, namely (a) Poisson, (b) exponential and (c) power-law. The right column panels (d, e, f) are sorted out by the infection status of individuals, namely (d) susceptible, (e) infected and (f) vaccinated. In each panel, the solid lines represent the theoretical results while the symbols are simulation results obtained by averaging over 100 realizations. Parameters are $N=900, D=30, v=0.1, m=10, r\in\{1,2,\dots,10\}, \langle r\rangle=3, \beta=0.005, \lambda=0.09, \varphi=0.005, \theta_0=0.1$.}\label{fig8}
\end{figure}

Figure \ref{fig8} displays the dependence of the final densities of individuals in each class on the tunable parameter $\alpha$ that reflects the vaccination strength relevant to the individual's effective interaction radius (\ref{func0}). In the left column panels the results are sorted out according to different distributions of interaction radius: (a) Poisson, (b) exponential and (c) power-law with the same average radius $\langle r\rangle=3$. Conversely, in the right column panels the same results are sorted out according to different disease status: (d) susceptible, (e) infected and (f) vaccinated. It is shown that the final density of susceptible (vaccinated) individuals reaches the minimum (maximum) at about $\alpha=0$, without respect to the radius distribution (see Figs.~\ref{fig8}(d,f)). In the case of $\alpha<0$, the vaccination favors susceptible individuals with small interaction radius. Moreover, the smaller the parameter $\alpha$, the larger the vaccination probability for susceptible individuals with smaller radius. Consequently, the final density of susceptible (vaccinated) individuals increases (decreases) as $\alpha$ decays since the vaccination covers only the susceptible individuals whose interaction radius is relatively small. As $\alpha$ becomes small enough, the vaccination only covers the susceptible individuals with the smallest interaction radius. Therefore the final densities of susceptibles and vaccinateds remain constant when $\alpha$ is extremely small. In the case of $\alpha>0$, the vaccination favors susceptible individuals with large interaction radius. The larger the value of $\alpha$, the higher vaccination probability for susceptible individuals with larger interaction radius. This implies that the final density of susceptible (vaccinated) individuals increases (decreases) with the increase of $\alpha$ since the vaccination takes place only for the susceptible individuals whose interaction radius is relatively large. However, when $\alpha$ is big enough, the vaccination only covers the susceptible individuals with the biggest interaction radius. Hence the final density of susceptible (vaccinated) individuals remains almost constant as $\alpha$ is extremely large. In general, as the value of $\alpha$ grows gradually, the final infected density first keeps almost unchanged, then drops to zero and remains in the disease-free state for a range of $\alpha$ with $\mathcal{R}_0<1$, and then rises fast and finally grows very slowly. The differences of the final density of each class of individuals between different radius distributions have also been demonstrated in Figs.~\ref{fig8}(d,e,f). In particular, as the value of $\alpha$ is large enough or small enough, the final density of infected individuals for the power-law radius distribution is lower than the infected density in the exponential radius distribution, which in turn is lower than that in the Poisson radius distribution (see Fig.~\ref{fig8}(e)). Otherwise, the interval of $\alpha$ that satisfies $\mathcal{R}_0<1$ for the Poisson radius distribution is wider than that in the exponential and power-law radius distributions. All the above results illustrate a good agreement between stochastic simulations and theoretical calculations, additionally the extreme points at about $\alpha=0$ may suggest an optimal vaccination intervention for disease control, regardless of the radius distribution.

\begin{figure}[htp]
\centering
\includegraphics[scale=0.4]{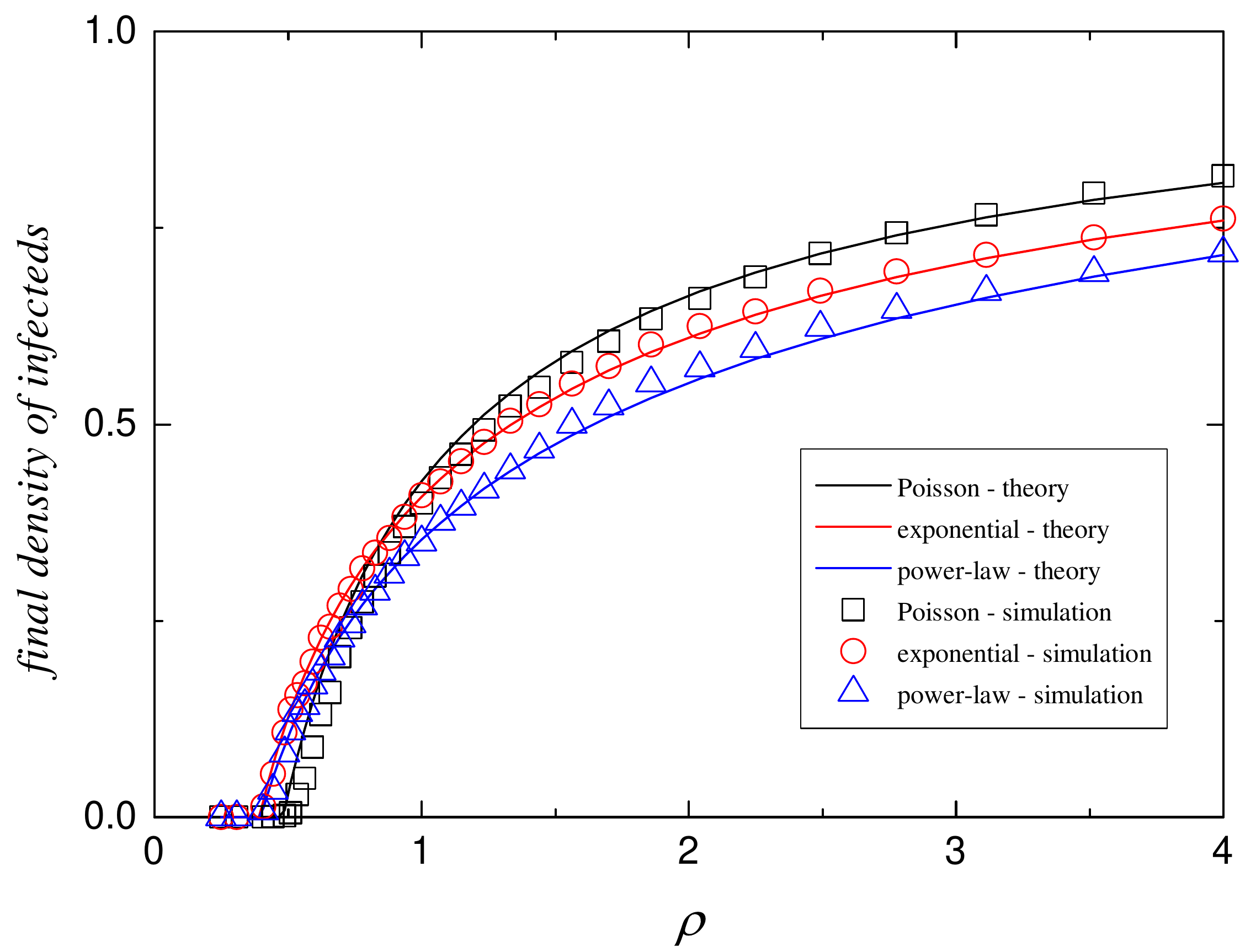}
\caption{(Color online) The final density of infected individuals as a function of individual density $\rho$ for different radius distributions with the same average $\langle r\rangle=3$. Symbols represent stochastic simulations averaged over 100 independent realizations and the solid lines stand for theoretical results. Parameters are $N=900, v=0.1, m=10, r\in\{1,2,\dots,10\}, p_{\rm jump}=1, \beta=0.005, \lambda=0.05, \varphi=0.005, \theta_0=0.1, \alpha=8$.}\label{fig9}
\end{figure}

We also extract the final epidemic prevalence as a function of the density $\rho$ of moving individuals. Results are reported in Fig.~\ref{fig9} where a comparison is exhibited among different distributions of interaction radius with the same expectation $\langle r\rangle=3$ and the same parameters of infection. It is observed that given all other parameters there exists a critical value of density $\rho_{\rm c}$, above which the disease breaks out, otherwise it dies out. In fact it can be deduced from Eq.~(\ref{func9}) that the threshold condition $\rho=\rho_{\rm c}$ is equivalent to $\mathcal{R}_0=1$, and that $\rho>\rho_{\rm c}$~$(\rho<\rho_{\rm c})$ is equivalent to $\mathcal{R}_0>1$~$(\mathcal{R}_0<1)$. On the other hand, when $\rho>\rho_{\rm c}$ the epidemic prevalence versus $\rho$ is monotonic, as intuition suggests that individuals in a population with higher density are more connected to each other, leading to a higher level of infection. In addition, it is shown that the value of $\rho_{\rm c}$ under the power-law and exponential radius distributions is smaller than that with Poisson radius distribution. As $\rho$ is far larger than $\rho_{\rm c}$ the epidemic prevalence in the population with Poisson radius distribution is larger than that with exponential radius distribution, which is in turn higher than that in the population with power-law radius distribution. This result restates the conclusion drawn from Fig.~\ref{fig7} that the disease inclines to spread in the population with more heterogeneous interaction radius; however, resulting in a relatively lower epidemic prevalence.

\begin{figure}[htp]
\centering
\includegraphics[scale=0.4]{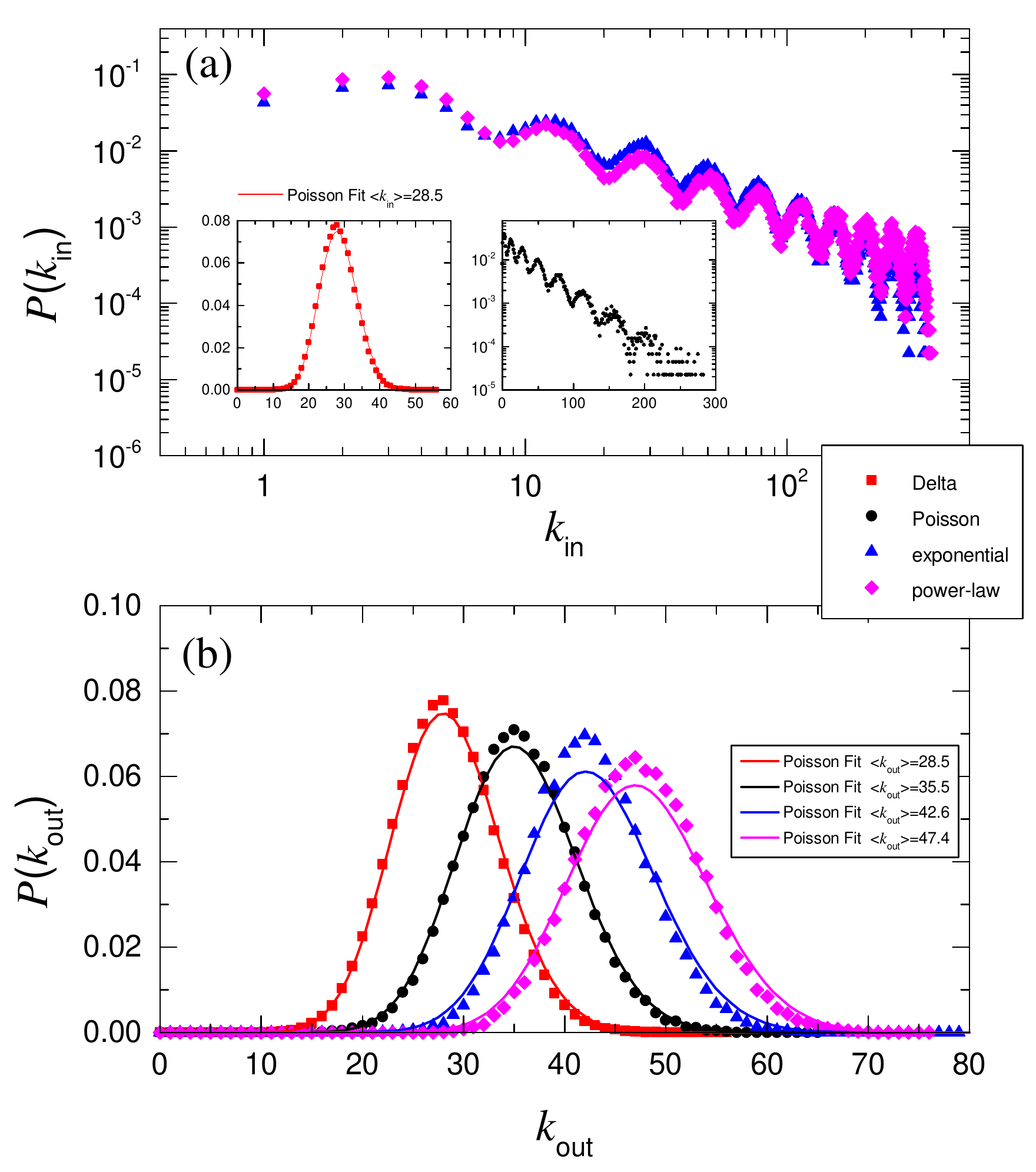}
\caption{(Color online) The probability distributions of (a) in-degree $P(k_{\rm in})$ and (b) out-degree $P(k_{\rm out})$ in the steady state with different radius distributions with the same average $\langle r\rangle=3$. The insets in panel (a) are results of $P(k_{\rm in})$ under the Delta (in linear-linear plot) and Poisson (in linear-log plot) radius distributions. Symbols represent stochastic simulations and the solid lines stand for the Poisson fittings with different expectations. Parameters are $N=900, D=30, v=0.1, m=10, r\in\{1,2,\dots,10\}, p_{\rm jump}=1$.}\label{fig10}
\end{figure}

\subsection{Structure of the directed contact network}

Figure~\ref{fig10} presents both the in-degree distribution $P(k_{\rm in})$ and the out-degree distribution $P(k_{\rm out})$ of the resulting directed contact network in the steady state for different interaction radius distributions of individuals. As shown in Fig.~\ref{fig10}(a), for moving individuals with power-law (marked by pink diamonds) or exponential (marked by blue triangles) radius distributions, the in-degrees of the directed contact network follow a multimodal distribution with a power-law decaying trend as shown in the log-log plot. In the case of Kronecker Delta radius distribution (marked by red squares), the in-degrees of the directed contact network follows a Poisson distribution with an expectation of $\langle k_{\rm in}\rangle=28.5$ (see the left inset in Fig.~\ref{fig10}(a)), while in the case of Poisson radius distribution (marked by black circles), the in-degree of the directed contact network obeys a multimodal distribution with an exponential decaying trend as shown in the semilog plot (see the right inset in Fig.~\ref{fig10}(a)). On the contrary, Fig.~\ref{fig10}(b) indicates that the out-degrees of the directed contact network follow a Poisson distribution, in which the expectation $\langle k_{\rm out}\rangle$ is relevant to the radius distribution. As illustrated by the Poisson fitting lines in Fig.~\ref{fig10}(b), the average out-degree $\langle k_{\rm out}\rangle$ of the directed network is $28.5, 35.5, 42.6, 47.4$ for Delta, Poisson, exponential and power-law distributions of interaction radius, respectively. The more heterogeneous the interaction radius distribution, the larger the average out-degree of the contact network.

\begin{figure}[htp]
\centering
\includegraphics[scale=0.4]{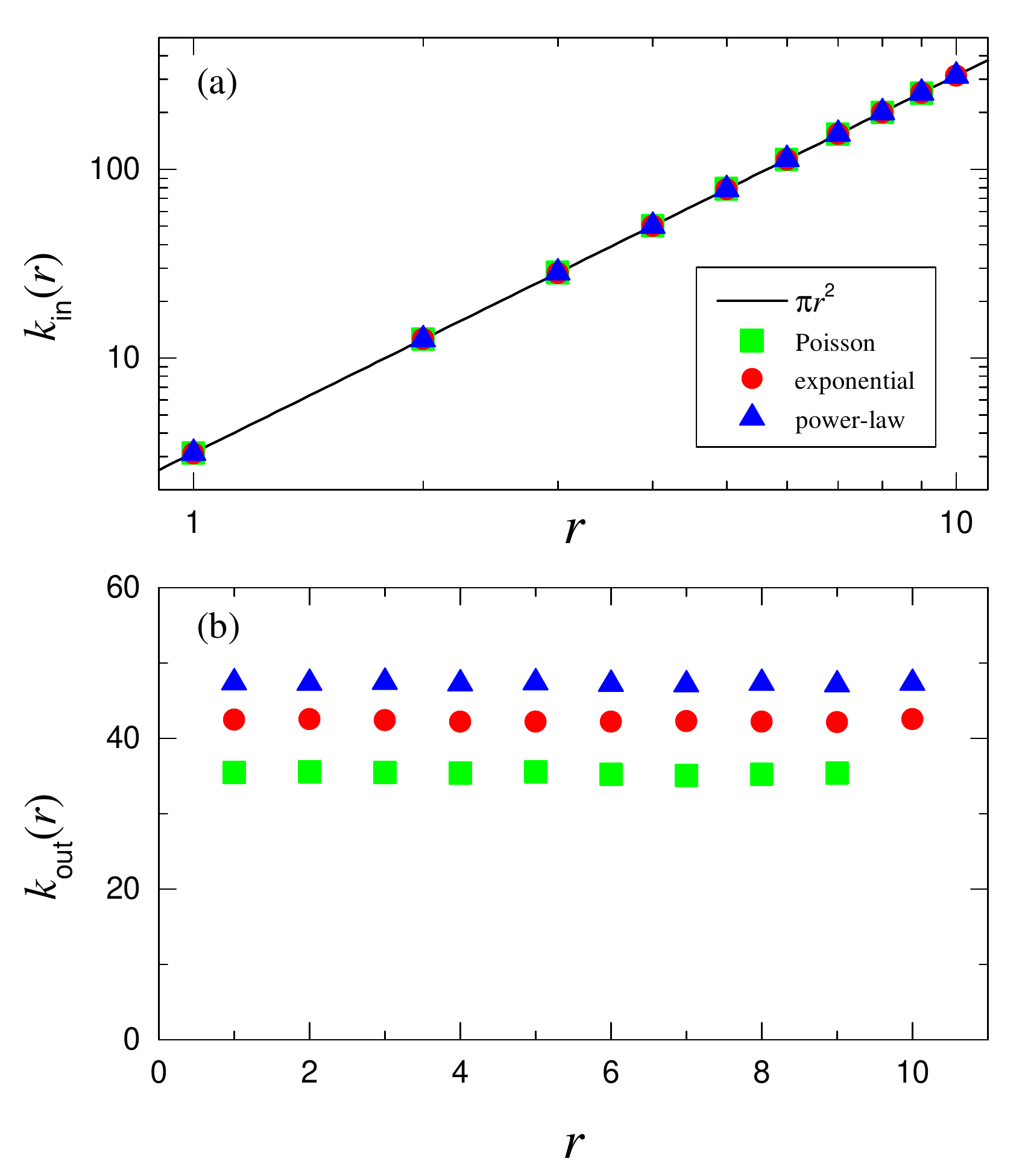}
\caption{(Color online) (a) The average in-degree $k_{\rm in}(r)$ and (b) the average out-degree $k_{\rm out}(r)$ of the individuals with radius $r$ as a function of $r$ for different radius distributions with the same average radius $\langle r\rangle=3$. Note that in the case of Poisson radius distribution, the result for $r=10$ is missing because the probability for an individual to have radius $r=10$ is vanishingly small. Parameters are $N=900, D=30, v=0.1, m=10, r\in\{1,2,\dots,10\}, p_{\rm jump}=1$.}\label{fig11}
\end{figure}

In Fig.~\ref{fig11} we provide the average in-degree $k_{\rm in}(r)$ and out-degree $k_{\rm in}(r)$ of individuals that have radius $r$ as a function of the radius $r$. It is clear from Fig.~\ref{fig11}(a) that $k_{\rm in}(r)$ is proportional to the area $\pi r^2$ of the effective contact space (circle) of individuals with radius $r$. This relation is not surprising and it is reasonable since the in-degree of an individual is defined as the total number of other individuals that fall within the circle with respect to the radius of the individual considered. Therefore, the larger the area of the contact circle, the larger the in-degree of the individuals. In this case, note that the density is $\rho=N/D^2=1$, we have exactly $k_{\rm in}(r)=\pi r^2$, as indicated by the black solid line in Fig.~\ref{fig11}(a). Moreover, it is shown that this result is independent on the radius distribution. In a sharp contrast, it is observed from Fig.~\ref{fig11}(b) that the average out-degree $k_{\rm out}(r)$ of individuals with radius $r$ is completely uncorrelated with the radius $r$. However, the value of the average out-degree is strongly related to the radius distribution. As demonstrated in Fig.~\ref{fig11}(b), we have $k_{\rm out}(r)=35.5$, $42.6$ and $47.4$ for Poisson, exponential and power-law distributions of interaction radius, respectively.

\begin{figure}[htp]
\centering
\includegraphics[scale=0.4]{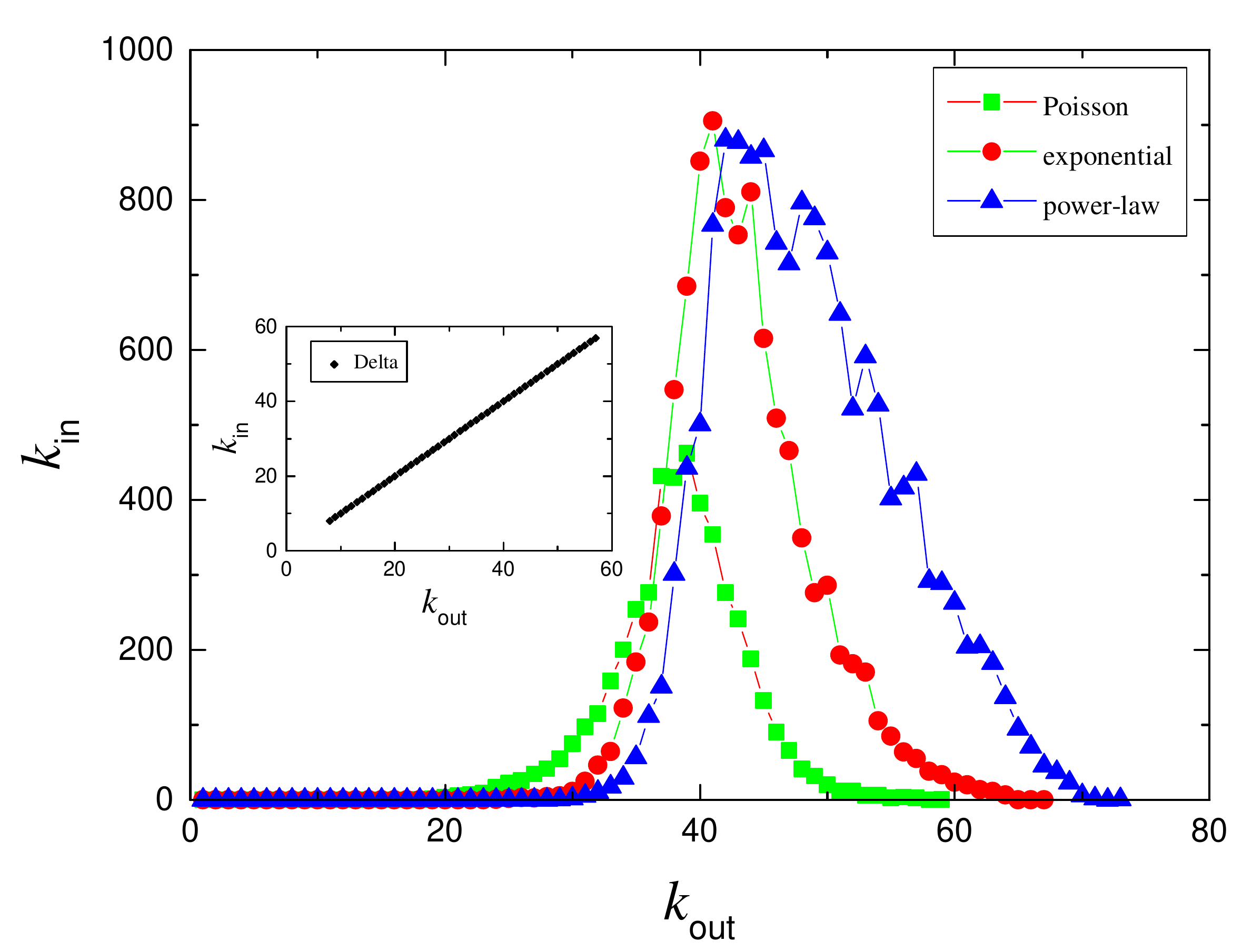}
\caption{(Color online) The correlation between in-degree $k_{\rm in}$ and out-degree $k_{\rm out}$ for different radius distributions with the same average radius $\langle r\rangle=3$. Parameters are $N=900, D=30, v=0.1, m=10, r\in\{1,2,\dots,10\}, p_{\rm jump}=1$.}\label{fig12}
\end{figure}

Figure~\ref{fig12} outlines the correlation between in-degree $k_{\rm in}$ and out-degree $k_{\rm out}$ for different cases of radius distributions. This correlation is determined by calculating the average in-degree of individuals whose out-degree is $k_{\rm out}$. In the cases of Poisson, exponential and power-law radius distributions, the correlation between $k_{\rm in}$ and $k_{\rm out}$ takes the unimodal form. As the out-degree increases, the in-degree first increases towards a summit and then decreases gradually to zero. In the case of the Delta radius distribution, we observe a linear correlation with $k_{\rm in}$ being equal to $k_{\rm out}$.

\begin{figure}[htp]
\centering
\includegraphics[scale=0.5]{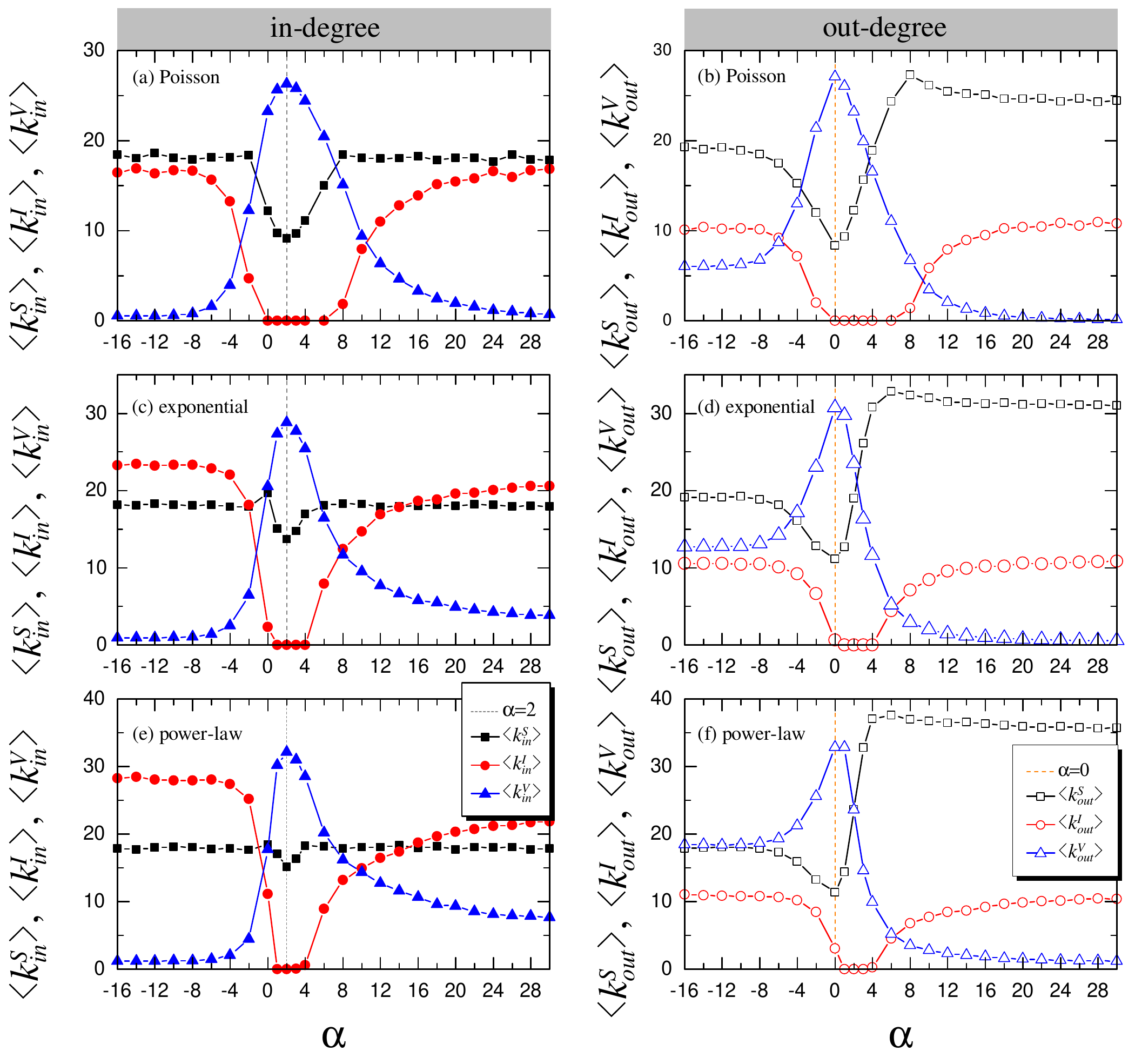}
\caption{(Color online) The average in-degrees of susceptible ($\langle k_{\rm in}^S\rangle$), infected ($\langle k_{\rm in}^I\rangle$) and vaccinated ($\langle k_{\rm in}^V\rangle$) individuals (a,c,e) and the average out-degrees of susceptible ($\langle k_{\rm out}^S\rangle$), infected ($\langle k_{\rm out}^I\rangle$) and vaccinated ($\langle k_{\rm out}^V\rangle$) individuals (b,d,f) as a function of $\alpha$. The gray (orange) dashed line is an auxiliary line for $\alpha=2$ (respectively, $\alpha=0$) to show the extreme points. Parameters are $N=900, D=30, v=0.1, m=10, r\in\{1,2,\dots,10\}, p_{\rm jump}=1, \beta=0.005, \lambda=0.09, \varphi=0.005, \theta_0=0.1$.}\label{fig13}
\end{figure}

In order to deep understand how the vaccination behavior affects the network structure and the transmission process, in Fig.~\ref{fig13} the average in-degree and average out-degree of individuals in each class are plotted against the vaccination-strength-related parameter $\alpha$ for different distributions of interaction radius. It is shown from the Figs.~\ref{fig13}(a,c,e) that the average in-degrees $\langle k_{\rm in}^S\rangle$, $\langle k_{\rm in}^I\rangle$ and $\langle k_{\rm in}^V\rangle$ of susceptible, infected and vaccinated individuals reached their respective extreme values at $\alpha=2$, irrespective of the radius distribution. A direct comparison with Fig.~\ref{fig4} suggests that the average in-degree of individuals is implicitly relevant to the behavior of the basic reproduction number $\mathcal{R}_0$ which arrives at the minimal value at $\alpha=2$.  On the other hand, Figs.~\ref{fig13}(b,d,f) show that the behavior of the average out-degree of individuals in each class closely resembles the final density of each class of individuals (see Fig.~\ref{fig8}) in that they reach their extreme values at about $\alpha=0$ without regard to the radius distribution. This similarity indicates that the average out-degree of individuals is potentially responsible for the behavior of the final density with regard to $\alpha$.

\section{Conclusions} \label{section7}

In summary, we have established an SIS epidemic model with vaccination on a dynamic network of mobile individuals with spatial constraints where all individuals have heterogeneous interaction radii (or contact circles) in a two-dimensional space. In the model, we consider that the vaccination of each susceptible individual depends on the individual's interaction radius and assume that all individuals are random walkers who are also allowed to perform a long-distance jump with a probability $p_{\rm jump}$. We derive a homogeneous mixing model with a set of ordinary differential equations in the special case of $p_{\rm jump}=1$. We have obtained the basic production number $\mathcal{R}_0$ and studied the dynamical behavior of the model. We argue that the disease-free equilibrium is locally asymptotically stable as $\mathcal{R}_0<1$; otherwise, if $\mathcal{R}_0>1$ then the disease-free equilibrium is unstable and there is a unique endemic equilibrium which is locally asymptotically stable and persistent. Extensive computational simulations have been carried out to further explore the dynamical behavior of the epidemic spreading as well as the topological properties of the underlying contact network. Our results have demonstrated a good agreement between theory and simulations with regard to the disease transmission. It is found that different distributions of individuals' interaction radius have strong impacts on the basic reproduction number and the final densities of individuals of each class. Generally speaking, the heterogeneity of interaction radius among individuals will facilitate the disease transmission while resulting in a relatively low epidemic prevalence. Moreover, based on the dependencies of $\mathcal{R}_0$ and the final epidemic prevalence on the vaccination-strength-related parameter $\alpha$, we argue that an optimal vaccination intervention is achievable for disease prevention and control. Furthermore, it is shown that the in-degree distribution of the resulting network in the cases of power-law and exponential radius distributions follows a multimodal distribution with a power-law decaying trend, whereas in the case of Poisson radius distribution, the in-degrees of the contact network follows a multimodal distribution with an exponential decaying trend. Conversely, in the case of Delta radius distribution, the in-degree distribution is Poisson. On the other hand, the out-degrees of the contact network follows a Poisson distribution with different expectations for different radius distributions. The correlation between in-degree and out-degree of the contact network takes the form of a unimodal function. It is worth noticing that the average in-degree and the average out-degree of individuals in each class have the same qualitative behavior with respect to the parameter $\alpha$. This resemblance provides another perspective to understand the influence of the vaccination intervention on the epidemic transmission as it has great impacts on the average in-degree and out-degree of individuals. This work provides an in-depth analysis for the dynamical behavior of the epidemic model and sheds new light on potentially optimal vaccination interventions for epidemic spreading in the population of moving individuals with spatial limitations.

\section*{Acknowledgments}

This research project was jointly supported by the National Natural Science Foundation of China (Grant Nos. 11601294, 61873154, 11501340, 11331009), Shanxi Province Science Foundation for Youths (Grant No. 201601D021012) and Shanxi Scholarship Council of China (Grant No. 2016-011).

\section*{References}

\bibliography{mybibfile}

\end{document}